\documentclass[letterpaper,aps,prd,preprint,showpacs,nofootinbib,superscriptaddress]{revtex4-1}
\usepackage{amsmath,amssymb,graphicx}
\usepackage[stable]{footmisc}
\usepackage{slashed}

\pdfoutput=1

\newcommand{\nc}{\newcommand}
\nc{\postscript}[2] 
{\setlength{\epsfxsize}{#2\hsize}\centerline{\epsfbox{#1}}}
\nc{\non}{\nonumber}
\nc{\hc}{\hbox {h.c.}} \nc{\re}{\hbox {Re}} \def\im{{\rm Im}}
\nc{\mev}{\hbox {MeV}} \nc{\gev}{\;\hbox {GeV}} \nc{\tev}{\;\hbox {TeV}}
\def\lsim{\mathrel{\raise.3ex\hbox{$<$\kern-.75em\lower1ex\hbox{$\sim$}}}}
\def\gsim{\mathrel{\raise.3ex\hbox{$>$\kern-.75em\lower1ex\hbox{$\sim$}}}}

\nc{\etal}{{\it et al.}}
\nc{\Lsp}{\;\;\;\;\;\;\;\;\;\;}  \nc{\LLLsp}{\lspace \lspace}
\nc{\lsp}{\;\;\;\;\;\;}
\nc{\spac}{\;\;\;}
\nc{\noi}{\noindent}
\nc{\beq}{\begin{equation}}   \nc{\eeq}{\end{equation}}
\nc{\bea}{\begin{eqnarray}}   \nc{\eea}{\end{eqnarray}}
\nc{\baa}{\begin{array}}      \nc{\eaa}{\end{array}}
\nc{\bit}{\begin{itemize}}    \nc{\eit}{\end{itemize}}
\nc{\ben}{\begin{enumerate}}  \nc{\een}{\end{enumerate}}
\nc{\bce}{\begin{center}}     \nc{\ece}{\end{center}}

\def\ie{{\it i.e.}}

\def\sq2{\sqrt{2}}

\def\ph{\varphi}

\def\m4{m^4(\ph)}
\def\mn2{m_n^2}

\def\v5{V^{(5)}}

\def\baa{\begin{array}}
\def\eaa{\end{array}}

\begin{document}

\begin{flushright}
 \mbox{\normalsize \rm CUMQ/HEP 181}\\
\end{flushright}

\vskip 20pt

\title{Higgs boson production and decay in 5D warped models}

\author{Mariana Frank\footnote{mariana.frank@concordia.ca}}
\affiliation{Department of Physics, Concordia University\\
7141 Sherbrooke St. West, Montreal, Quebec,\\ CANADA H4B 1R6
}
\author{
Nima Pourtolami\footnote{n\_pour@live.concordia.ca}}
\affiliation{Department of Physics, Concordia University\\
7141 Sherbrooke St. West, Montreal, Quebec,\\ CANADA H4B 1R6
}
\author{Manuel Toharia\footnote{mtoharia@dawsoncollege.qc.ca}}
\affiliation{Department of Physics, Concordia University\\
7141 Sherbrooke St. West, Montreal, Quebec,\\ CANADA H4B 1R6
}
\affiliation{Physics Department, Dawson College\\
 3040 Sherbrooke St., Westmount, Quebec,\\ CANADA H3Z 1A4
}

\date{\today}

\begin{abstract}

We calculate the production and decay rates of the Higgs boson at the LHC in the
context of general 5 dimensional (5D) warped scenarios with a spacetime background
modified from the usual $AdS_5$, with SM fields propagating in the
bulk. We extend previous  work  by considering the full
flavor structure of the SM, and thus including all possible flavor
effects coming from mixings with heavy fermions. We proceed in three different ways,
first by only including two complete Kaluza-Klein (KK) levels ($15\times15$ fermion mass matrices),
then including three complete KK levels ($21\times21$ fermion mass matrices) and
finally we compare with the effect of including the infinite (full) KK towers. We present
numerical results for the Higgs production cross section via gluon
fusion and Higgs decay branching fractions in both the modified
metric scenario and in the usual  Randall-Sundrum metric scenario.

\end{abstract}

\pacs{11.10.Kk, 12.60.Fr, 14.80.Ec}

\maketitle

\section{Introduction}
\label{sec:intro}

The discovery of a light Higgs-like boson at the first run of the LHC
seems to have provided an answer to the question of the origin of particle
masses. While the particle discovered resembles very much the Standard
Model (SM) Higgs boson, variations from the predicted coupling
strengths of the SM are possible \cite{Aad:2015gba,Chatrchyan:2012xdj}.

These variations would be related to models beyond the SM, which
address some of the shortcomings of the current theory such as the
hierarchy problem or the flavor puzzle.
Unfortunately the LHC has not yet provided an unequivocal
signal for physics beyond the SM (supersymmetry, extra-dimensions)
and such new physics scenarios could manifest themselves in some spectacular signal to be yet
detected at Run II of the LHC. But they could also arise from a precise and careful measurement of
the properties of the newly discovered 
boson, such as its mass, its
couplings, its width or its production and decay rates.


In this work  we investigate the signal strengths for gluon fusion
production as well as tree-level and loop-dependent couplings for a broad class of models
in which the space-time is extended to a warped geometry model with a
five-dimensional background space-time metric. The initial incentive
for these models was to solve the weak-Planck scale hierarchy by
allowing gravity to propagate in the 
bulk of the extra dimension \cite{Randall:1999vf} which has to be 
stabilized \cite{stabilization}. Later it was
realized that by  allowing the SM fermion fields to propagate into the
bulk, different geographical localization of fields along the extra
dimension could help explain the observed masses and flavor mixing
among quarks and leptons \cite{Davoudiasl:1999tf, Pomarol:1999ad,  Grossman:1999ra,
  Chang:1999nh, Gherghetta:2000qt,   Davoudiasl:2000wi}. Electroweak
symmetry breaking can still happen via a standard Higgs mechanism in 
these scenarios (although the Higgs can also be implemented as as
pseudo-Nambu-Goldstone boson) \cite{Contino:2003ve,Csaki:2008zd}. The Higgs boson
itself must be located near the TeV boundary of the extra dimension in
order to solve the hierarchy problem, and so typically it is assumed
to be exactly  localized on that boundary (brane Higgs
scenario). Nevertheless, it is possible that it leaks out into the
bulk (bulk Higgs scenario), and in doing so, indirectly it can alleviate some
of the flavor bounds and precision electroweak tests plaguing these models
\cite{Carena:2004zn, Huber:2003tu, Agashe:2004cp, Agashe:2006at,
  Csaki:2008qq}. In order to satisfy the current bounds from precision flavor and
electroweak processes, and still have light enough new physics to be
seen at the LHC, one viable alternative is to extend the gauge group
\cite{Agashe:2003zs,Agashe:2006at}. Another possibility is to modify the
warping of the space-time metric, which can also
alleviate some of the more stringent bounds
\cite{Falkowski,Batell,Cabrer:2011fb,Carmona:2011ib,MertAybat:2009mk}. In
this work we focus on this last option. 

In a previous work, we investigated Higgs boson production  when
 allowed to propagate in the bulk both using the original Randall-Sundrum (RS) metric
\cite{Frank:2013un}, and also within a modified metric background
\cite{Frank:2013qma}.
We analyzed the Higgs production rate via gluon fusion
 and showed that the results are consistent with the LHC Higgs
 measurements, in the same region of the parameter space where flavor
 and precision electroweak constraints are safe.

However, in both of these instances, our analyses employed
a toy-model setup, in which the Higgs field was allowed to propagate
in the bulk accompanied by a single 5D fermion field. Here, we extend
our results to a realistic model with three families and include the full
flavor effects.
In addition to production cross section, we also analyze the Higgs couplings to quarks and
leptons as well as the branching ratio for the di-photon decay. We
include results from analyzing the model with 
a cut-off scale, and including two or three KK fermion levels only. We
then compare these results with those obtained including the complete
(infinite) tower of KK modes.
We present the results for the model with the modified metric, $MAdS_5$, as well as the results 
within the original metric, whose effects on Higgs physics with full
flavor was also studied in \cite{Archer:2014jca}.

Our work is organized as follows. In Sec. \ref{sec:1} we introduce our
model and discuss the limits imposed on its parameter space from precision
measurements. In Sec. \ref{sec:2} we analyze the effects of the KK modes on gluon fusion production of the
Higgs boson, assuming a finite number of KK modes (2 KK levels, then 3
KK levels). The same investigation for the $h \to \gamma \gamma$ coupling is presented in
Sec. \ref{sec:3}. Sec. \ref{sec:4} presents a detailed analysis of the
procedure involved in calculating the inclusion of the infinite tower of
KK fermions including three families of fermions. We then summarize
our findings and conclude in Sec. \ref{sec:summary}.  

\section{The $MAdS_5$ Model}
\label{sec:1}

Models with general warped extra space dimensions  ($MAdS_5$) are characterized by
the following metric \cite{Randall:1999vf} 
\beq
ds^2 = e^{-2 A(y)} \eta_{\mu\nu}dx^{\mu}dx^{\nu} + dy^2,\label{metricansatz}
\eeq
where $\eta_{\mu\nu} = \text{diag}(-1,1,1,1) $ and $A(y)$ is a function of the extra space dimension 
originally (in RS models) assumed to be
\bea
A(y) = k y,
\eea
with $k$ being the inverse curvature radius of the $AdS_5$ space-time.
In these models the extra dimension, $y$, is bounded by two branes (hard walls) located at 
$y=0$ and $y=y_1$, corresponding to the UV and IR scales respectively. 
In  \cite{Cabrer:2011fb} it has been shown that, assuming the  superpotential to be
\bea
W_{\phi} (\phi) = 6 k (1+ b e^{\nu\phi/\sqrt{6}}),
\eea
with real parameters $b$ and $\nu$, modifies the background configuration
for the metric warp function, $A(y)$, and the scalar field, $\phi (y)$. The dilatonic scalar 
field in addition to the SM scalar Higgs field emerges as
\bea\label{genericmetric}
A(y) = ky - {1\over \nu^2} \text{log}\left(1-{y\over y_s}\right), \ \ \ \ 
\phi(y) = -{\sqrt{6}\over \nu} \text{log}\left[\nu^2 bk(y_s - y)\right],
\eea
where $y = y_s$ is the position of the singularity of the metric,  generated by the 
scalar field $\phi (y)$. 
For this geometry the curvature $kL$ and the curvature radius $R$ are modified from the $AdS_5$ case
and are given by
\bea\label{curvature}
kL(y) = { k \Delta \nu^2\over\sqrt{1-2\nu^2/5 + 2k\Delta\nu^2+(k\Delta)^2\nu^4}},
\eea
and
\bea
R(y) = -20 k^2{ (1 - 2/5 \nu^2 + 2 k\Delta\nu^2 +  (k\Delta)^2\nu^4\over (k\Delta)^2\nu^4},
\eea
respectively, where $k\Delta = k(y_s - y)$ is always positive, \ie,~ the singularity is always assumed to
be outside of the physical region. We refer to this scenario as the modified $AdS_5$ ($MAdS_5$) model. 
\begin{figure}[h]
\center
\begin{center}
	\includegraphics[height=6.cm]{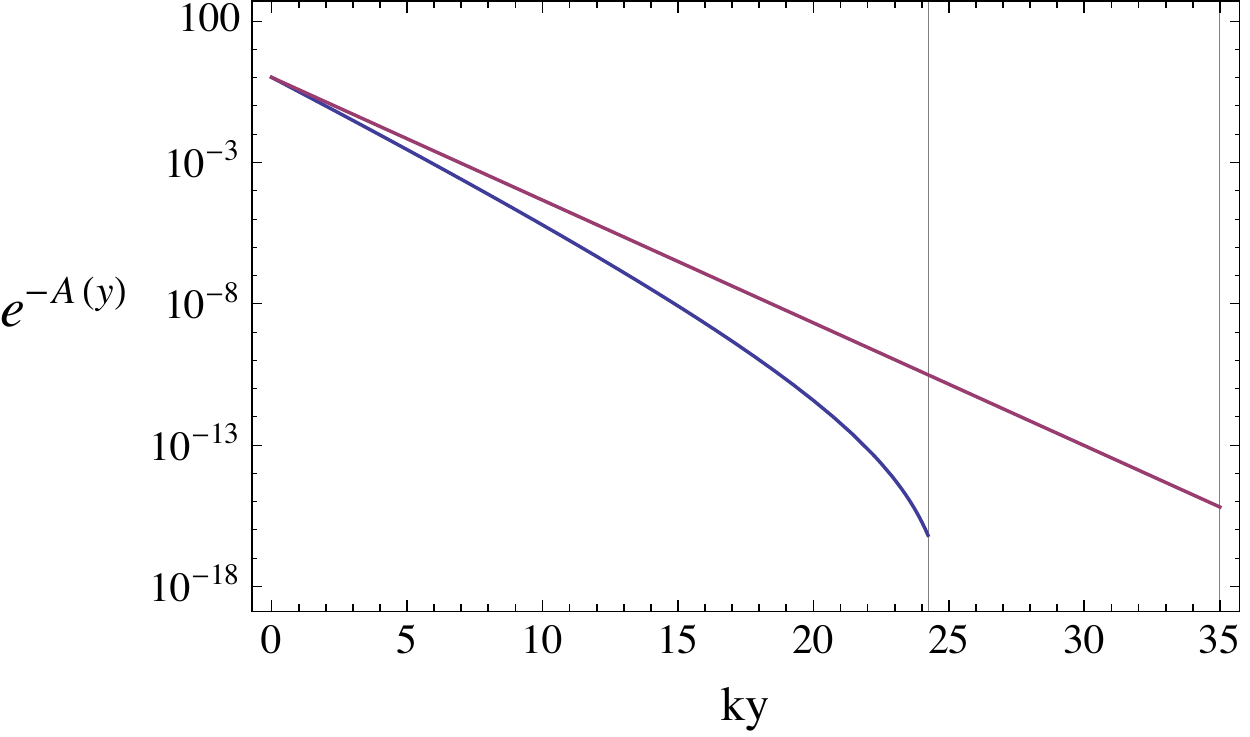}\hspace{-.3cm} 
 \end{center}
\caption{Value of the warp factor as a function of the length along
  the extra dimension (in units of $k$) for both the RS (linear plot
  in pink) and for the $MAdS_5$ (curved plot in blue) scenarios.  
 }  \label{fig:TA}
\end{figure}
In Fig. \ref{fig:TA} we show the value of the warp factor as a function of the length along
  the extra dimension (in units of $k$) for both the RS (in pink)
  and for the $MAdS_5$ (in blue) scenarios showing how, in the $MAdS_5$
  case, one produces the Planck-TeV hierarchy with a shorter
  extra-dimensional length due to stronger warping near the TeV boundary.  

The bulk Higgs mass is given by
\bea
M^2_H = a (a-4) k^2 \left(1-{4\over (a-4)k\Delta\nu^2}\right),
\eea
yields the following  equation of motion for the Higgs profile 
\bea
-\partial_y\left(e^{-4A(y)}\partial_y h(y)\right) + M^2_H e^{-4A(y)} h(y) -\lambda^2 e^{-2A(y)}h(y) = 0.
\eea
Here $a\in \mathbb{R}$ is the bulk mass parameter of the Higgs field. This parameter determines the
localization of the Higgs profile along the extra dimension. For values of $a\gtrsim 10$, the Higgs field is 
localized on the IR brane and it can effectively be described via a Dirac delta function (\cite{Frank:2013un,Azatov:2010pf}).
At the lower limit, $a_{min}$, the Higgs field will be as delocalized as possible while still 
offering a solution to the hierarchy problem, as explained below. 

Introducing the boundary condition as $(\partial_y -M_0) h(y)|_{UV} = 0$, the solution to the equation 
of motion for the Higgs profile can be written as 
\bea\label{higgsfull}
h(y)
=h_0 e^{aky} \bigg[1 + (M_0/k -a) \left[ F(y) - F(0) \right]\bigg], 
\eea
where $h_0$ is a normalization factor and 
$M_0$ is the brane Higgs mass term (the coefficient of the Higgs boundary potential $|H|^2\delta(y-y_1)$ at the
IR brane) introduced to give rise to the Higgs zero mode field with the correct physical mass. The function $F(y)$ is given by
\bea
F(y) =  e^{ - 2 (a-2) k y_s }k y_s \left[ -2(a-2) k y_s \right]^{-1 + 4/\nu^2} \Gamma \left[ 1 - \frac{4}{\nu^2} , -2(a-2) k( y_s - y) \right].
\eea
As seen from the Higgs profile, Eq. (\ref{higgsfull}), only the first term, $h = h_0 e^{aky}$ can address
the hierarchy problem of the SM. The second term, which is peaked at the UV brane, corresponding to an 
elementary Higgs field, must be subdominant in order to preserve localization of the Higgs field near the IR 
brane. One could fine tune $M_0/k\simeq a$,  but in order to
avoid this fine-tuning of the boundary mass term,  demanding that 
\bea\label{delta}
\delta \equiv |F(y_1)|\sim{\cal O}(1),
\eea
would be sufficient,  as $F(y)$ is a monotonically increasing function.
In the following we set $\delta \simeq 0.1 - 1$ to ensure that we do not introduce
a new fine-tuning to the setup.

Once the Higgs vacuum expectation value (VEV) profile is set, the lightest KK mode in the Higgs
sector will be identified as the SM Higgs boson, with a mass of 125
GeV. In general terms, one expects the mass of all KK modes to be of
similar order, i.e. TeV size. In order to generate a mass 10-20 times
lighter one must require some $1-10\%$ tuning of parameters in the
scalar sector in order to address this small hierarchy and one should
also include all possible quantum corrections to the tree level KK Higgs 
mass. As long as $\delta \simeq 0.1 - 1$, this small tuning in the
Higgs sector will just be a reflection of the remnant hierarchy between the TeV scale and the 
electroweak scale, far less worrisome than the Planck-electroweak hierarchy.

Furthermore, it is not obvious that
the lightest scalar field in the model should be the Higgs boson. After all, the
scenario makes use of a 5D scalar singlet, whose perturbations, mixed
with the graviscalar perturbations would generate also a tower of new
scalar fields. The lightest one of them, identified as the
radion/dilaton could be light, although the expectation is that the
strong deviation from $AdS_5$ should lift its mass to more generic
values of KK fields, i.e. TeV range. This was addressed in
\cite{Cabrer:2011fb} and found dilaton masses consistently growing
with the level of deviation from $AdS_5$. Finally, it is more than
possible that the dilaton tower could mix with the Higgs tower, so
that the lightest scalar field in the mixed scalar sector would be
some mixture of 5D Higgs, graviscalar and 5D scalar singlet. We will work under the
assumption that the graviscalar/5D-singlet tower mixes minimally with
the lightest Higgs, so that this last one remains Standard-Model-like
(as preferred by experimental data). The next and heavier scalar excitations
can still be highly mixed states inheriting Higgs-like, radion-like or
sterile couplings but any further study of this sector is beyond the scope of
this work. 

What makes the model presented in Eq. (\ref{genericmetric}) so interesting is the fact that
it substantially alleviates the bounds on the KK masses due to the electroweak
precision parameters \cite{Cabrer:2011fb}. The parameters that introduce the tightest bounds on the 
$M_{KK}$ are $T$ and $S$ parameters which are given by
\bea\label{TandS}
T &=& {1\over \alpha} s_W^2 m_Z^2 y_1 \int e^{2 A(y)} (\Omega_f - \Omega_h)^2\\
S &=& {8 \over \alpha} s_W^2 c_W^2 m_Z^2 y_1 \int e^{2 A(y)} (\Omega_f-y/y_1)(\Omega_f - \Omega_h),
\eea
where the functions $\Omega_s$ for the scalar and $\Omega_f$ for the fermion fields are given by
\bea\label{OmegaDef}
\Omega_{h}(y) = \int_0^{y}e^{-2A(y)}h^2(y)\\\nonumber
\Omega_{f}(y) = \int_0^{y}e^{-3A(y)}f^{2}(y),
\eea
with $h(y)$ and $f(y)$ the zero mode profiles of the scalar and fermion fields. The functions $\Omega$ have 
the property  that $\Omega_h(0) = 0$ and $\Omega_f(y_1) = 1$. For a UV localized
field (\ie ~light fermions and gravitons), $\Omega_f(y) = 1$. Substituting the zero mode profiles of the
fields \cite{Frank:2013qma}, we get
\bea\label{Omega}
\Omega_h(y) &=&  {\Gamma(1+2/\nu^2,2(a-1)ky_s) - \Gamma(1+2/\nu^2,2(a-1)k\Delta)\over\Gamma(1+2/\nu^2,2(a-1)ky_s) - \Gamma(1+2/\nu^2,2(a-1)k\Delta_1)}\\
\Omega_f(y) &=&  {\Gamma(1-(1-2c)/\nu^2,(1-2c)ky_s) - \Gamma(1-(1-2c)/\nu^2,(1-2c)k\Delta)\over\Gamma(1-(1-2c)/\nu^2,(1-2c)ky_s) - \Gamma(1-(1-2c)/\nu^2,(1-2c)k\Delta_1)}
\eea
where $k\Delta_1 \equiv k(y_s - y_1)$. 
Experimentally the $S$ and $T$ parameters are found \cite{Agashe:2014kda} 
\bea\label{TandSpdg}
S = 0.00^{+0.11}_{-0.10},\\
T = 0.02^{+0.11}_{-0.12}.
\eea

\begin{figure}[t]
\center
\begin{center}
	\includegraphics[height=9.cm]{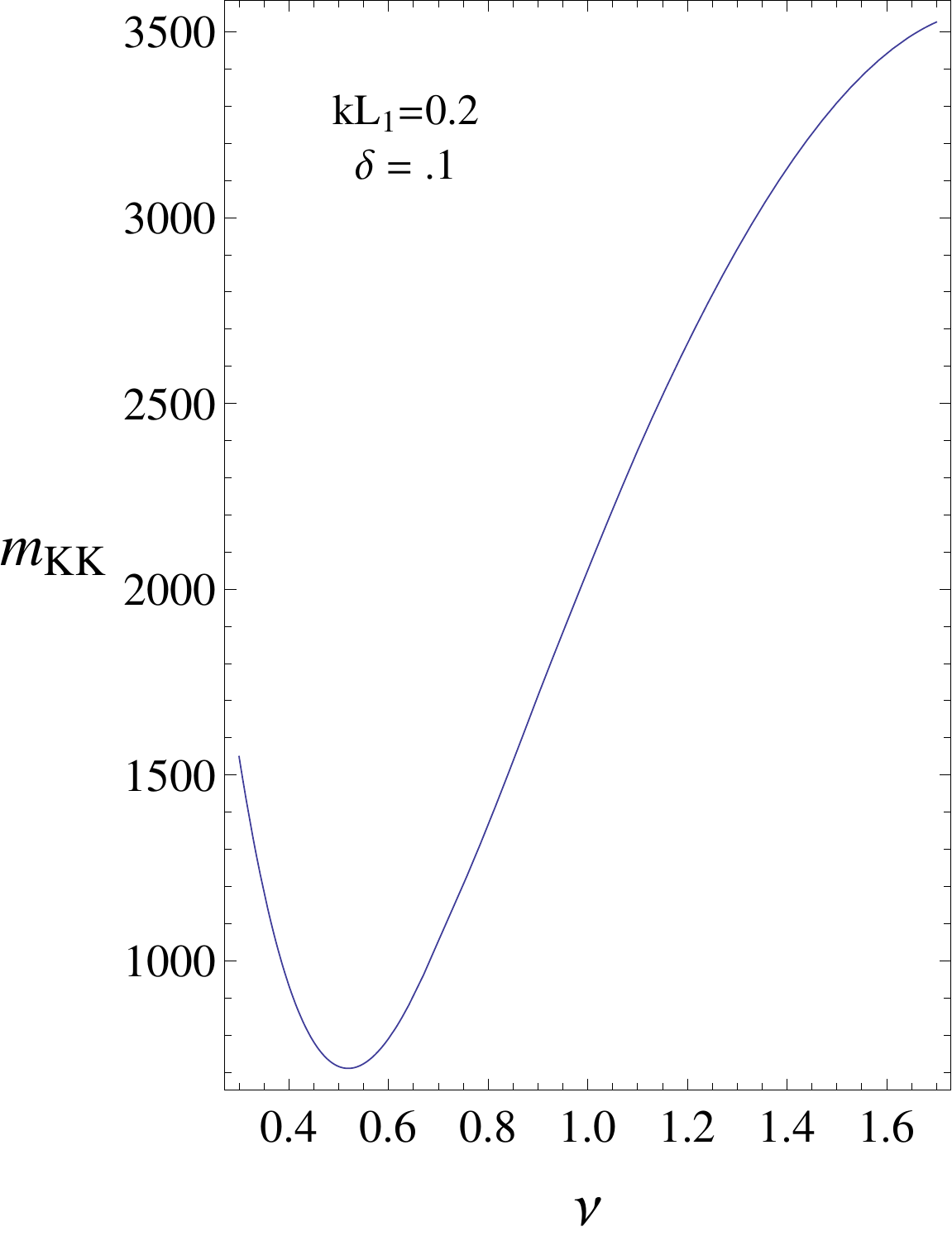} \hspace{.5cm}
	\includegraphics[height=9.cm]{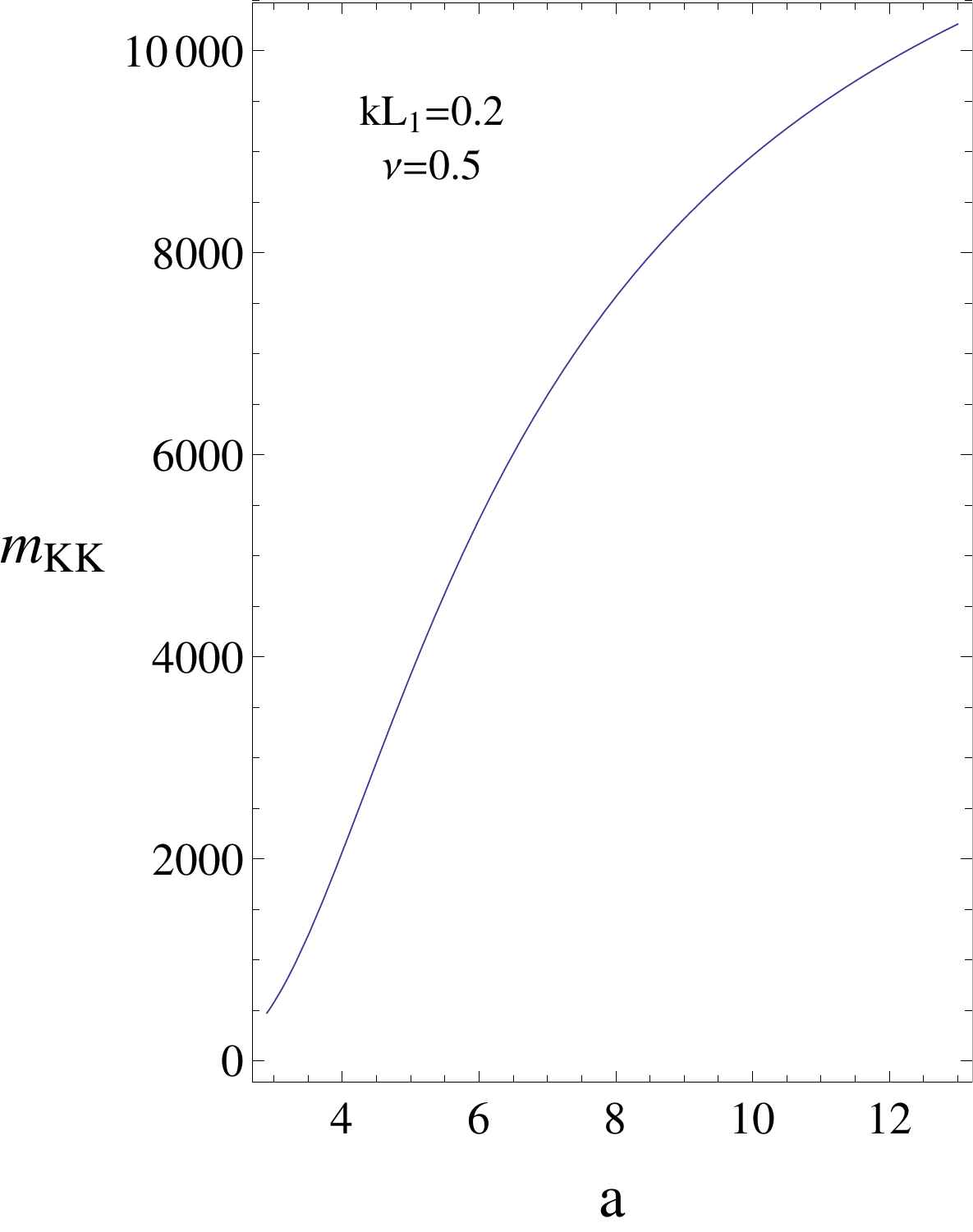} 	
 \end{center}
\caption{Electroweak precision test bounds on the KK masses as 
  functions of the modified metric parameter $\nu$ (left panel) and of the Higgs localization
  parameter $a$ (right panel), obtained by restricting $\Delta S=(-0.1, 0.11),~\Delta T=(-0.12, 0.11)$.
 }  \label{fig:mKKnu}
\end{figure}

In Fig. \ref{fig:mKKnu} we show the bounds that the $S$ and $T$ ranges impose on the  parameter
region (expressed as KK mass scale $m_{KK}$) of the $MAdS_5$ model, with the curvature radius in units of $k$, at the IR brane, $kL_1\equiv kL(y_1) = 0.2$. 
We have only considered the case for UV localized fermion interactions, where $\Omega_f = 1$. A
full analysis of the parameter space of this model is available in \cite{Cabrer:2011fb}. The left panel  of Fig. \ref{fig:mKKnu} shows the dependence of the mass scales with the $\nu$ parameter. 
In generating each point  we  fixed $kL_1 = 0.2$ and (for the right panel)  $\nu=0.5$, which in turn fixes 
the value for $k\Delta$, given by considering the positive solution to $kL(k\Delta,\nu) = kL_1$. With 
$k\Delta$ given,  we obtain the bulk Higgs parameter, $a$, and the position of the IR brane, 
$y_1$ through finding the simultaneous solutions to $\delta(y_1,\nu,k\Delta,a) = 0.1$ and for either
 the $T(y_1,\nu,k\Delta,a)$ or the $S(y_1,\nu,k\Delta,a)$ parameter which  satisfies the bounds given by Eq. (\ref{TandSpdg}).  

As it is evident from the figure, the scale of the IR brane (or $m_{KK}$) is determined by $A(y_1)$
which in turn is related to the volume of the extra dimension. The figure also shows that the values
of $kL_1 = 0.2$ and $\nu = 1/2$ result in best bounds on the KK masses, which can be as low as $\sim 700$ GeV. It turns out that after  considering
all possible values of $kL_1$, and scanning the parameter space of these models, this region still provides the 
most relaxed constraints on the KK masses. For this reason, in our numerical analysis provided in the 
next sections, we use this parameter space and contrast it to the RS regime,  given by taking the 
limits  $\nu \rightarrow\infty$ and $y_s\rightarrow\infty$. Note that the latter corresponds to bounds on $m_{KK}$ of order 10 TeV \cite{Huber:2000fh}.

It is also worth noting that the localization of the Higgs field 
has also an important effect on the  bounds shown in
Fig. \ref{fig:mKKnu}. The lowest value of $a$ that we consider is
$a_{min}$, defined as the value of the $a$-parameter such that $\delta
= 0.1$, and therefore such that any value of  $a \ge a_{min}$ will
require no fine-tuning in order to solve the gauge hierarchy 
problem. In the same figure, we show the relationship between the $a$
parameter and the lowest allowed KK mass, $m_{KK}$. As one can see
from the figure, higher $a$ parameter values introduce tighter bounds
on the KK masses from the electroweak precision test constraints.

\section{Fermion Yukawa Couplings and $gg \to h$  cross section}
\label{sec:2}
In this section we present the numerical calculations for the $hgg$ couplings considering
the full flavor structure of the SM. We present our results for two
cases, first for the choice $\{kL(y_1),\nu\} = \{0.2, 0.5\}$,
which corresponds to the parameter region where electroweak precision
constraints are the smallest, and then in the limit
$\{kL(y_1),\nu\} = \{0.9999, 10\}$ which corresponds to the RS
regime. In order to compare the two scenarios we set the value of
$A(y_1)$ in each case such that the lightest KK masses are about 
$2 - 2.5$ TeV. Having fixed $kL_1,~ \nu$ and $A(y_1)$, we solve for $y_1$ and 
$y_s$ from Eqs. (\ref{genericmetric}) and (\ref{curvature}). We also
obtain the value for $a_{min}$ (the lowest Higgs field localization mass
parameter) by requiring that $\delta(y_1,a) = 0.1$, see Eq. (\ref{delta}).
Due to the costly computational procedure, for
the numerical results in this paper we only consider $a$ such that 
$a \in \{a_{min}, a_{min}+0.5, a_{min}+1,\dots, a_{max} < 6.5\}$.
With the background metric parameters fixed, we then construct a fully realistic model 
which reproduces {\it all}  ~the SM masses and mixing angles. 
For this we scan over random anarchic fermion bulk mass parameters,
(\ie,  $c$-parameters)\footnote{For the $c$-parameters we guide the
  process by limiting the search near some fixed values that are known
  to produce correct masses and mixings to first approximation}  and the 5D
Yukawa couplings (we consider two situations, first the case where
the Yukawa couplings are of order $\sim 1$ and then when they are of order
$\sim 3$).
  
Using these values for $Y_{5D}$'s and $c$'s, and only considering the zero mode profiles 
for which analytic expressions are available\footnote{See \cite{Frank:2013qma} for full analytical 
expressions of the zero mode overlap integrals.}, we construct the
$3\times 3$ Yukawa coupling matrix with the following elements
\beq\label{YYY100}
(y^0_u)_{ij} = \frac{(Y^{5D}_u)_{ij}}{\sqrt{k}} \int_0^{y_1} dy e^{-4A(y)} h(y) q^{0,i}_L(y)u^{0,j}_R(y)\, ,
\eeq
where $ (Y^{5D}_u)_{ij} $ are the 5D dimensionless Yukawa couplings and 
$u \in \{u, d\}$ for up and down quarks, and $q^{0,i}_L(y)$ and
$u^{0,j}_R(y)$ are the $SU(2)$ doublet and singlet zero mode quark wave functions (the SM
quarks) in the gauge basis, obtained in the limit when the Higgs vacuum expectation value (VEV) $v\to 0$. 

When the previous Yukawa matrix reproduces well enough the SM masses
and mixings, we construct the first 2 KK (and later we repeat for
3 KK) profiles for fermions by solving numerically the differential
equations for the equations of motions of the fermion profiles 
for all  6 flavors (in the gauge basis and in the limit $v\to 0$):
\bea
\partial_y\left(e^{(2c-1)A(y)}\partial_y\left( e^{-(c+2)A(y)}\right)\right)f(y)+e^{(c-1)A(y)}\lambda^2 f(y) =0, 
\eea
with Dirichlet boundary conditions for the ``wrong'' chirality
fermions.
Equipped with the KK profiles,
we compute all Yukawa couplings in the setup with $N$ complete KK
levels, and construct the Yukawa matrix of dimension $(3+6N)$,
i.e. $15\times 15$ when the number of KK levels is $N=2$, and $21\times
21$ when the  KK levels are $N=3$. We write, for the up sector
\begin{equation}
{\bf Y}_{u} = \left(\begin{array}{ccc} 
(y^{0}_{u})_{3\times3}      &  (0)_{3\times 3N}   & (Y^{qU})_{3\times 3N}\\
 (Y^{Qu})_{3N \times3}      & (0)_{3N\times  3N} & (Y_1)_{3N \times 3N}\\
(0)_{3N \times3}      &     (Y_2)_{3N\times  3N} &  (0)_{3N\times 3N}
\end{array}\right),\label{yumatrix}
\end{equation}
with the down sector Yukawa matrix ${\bf Y}_d$ computed in the same
way. The submatrices are obtained by the overlap integrals 
\bea\label{YYY1KK}
Y^{qU}= \frac{Y^{5D}_{ij}}{\sqrt{k}} \int_0^{y_1} dy e^{-4A(y)} h(y) q^{0,i}_{L}(y)U^{n,j}_{R}(y) \\
Y^{Qu} = \frac{Y^{5D}_{ij}}{\sqrt{k}} \int_0^{y_1} dy e^{-4A(y)} h(y) Q^{m,i}_{L}(y)u^{0,j}_{R}(y)\\
Y_1 = \frac{Y^{5D}_{ij}}{\sqrt{k}} \int_0^{y_1} dy e^{-4A(y)} h(y) Q^{m,i}_{L}(y)U^{n,j}_{R}(y) \\
Y_2 = \frac{Y^{5D}_{ij}}{\sqrt{k}} \int_0^{y_1} dy e^{-4A(y)} h(y) Q^{m,i}_{R}(y)U^{n,j}_{L}(y)\, ,
\eea
where the indices $m$ and $n$ track the KK level and $i$ and $j$ are
flavor indices. 
The corresponding fermion mass matrix ${\bf M}_{u}$, of dimension $(3+6N)$, is given by 
\bea
{\bf M}_u = 
\left( \begin{matrix} vy^0_u & 0 & vY^{qU}\\
vY^{Qu} & M_Q & vY_1\\
0 & vY_2 & M_U
\end{matrix}\right) 
\label{mumatrix}
\eea
where $M_Q$ and $M_U$ are the $3N\times 3N$ diagonal KK mass matrices
and where $v = 174 $ GeV. We construct a similar mass matrix ${\bf M}_d$ for the down sector. 

We now have to redefine fields to go to the mass basis by diagonalizing ${\bf M}_{u}$
through the bi-unitary transformation 
\bea
{\bf M}_u \rightarrow V_L {\bf M}_u V_R.
\eea
The same transformation has to be applied to the Yukawa matrix as well
\bea
 {\bf Y}_u \rightarrow V_L {\bf  Y}_u V_R.
\eea
Note that the transformed $ {\bf Y}^{phys}_u$ matrix by this
procedure is not necessarily diagonal. Also note that at this point, due to the contribution of 
the tower of KK modes, the naive 0-mode masses and CKM mixings which where
arrived at through our first scan, might have been significantly shifted. This amount of this shift 
depends on the value of the $a$-parameter, and is most significant for the top quark mass. 
Following a try and error method, we go back to the first step and redo the scanning and 
refine our screening to find the $c$-parameters and 5D Yukawa
couplings such that these final masses and CKM
mixing angles, which include mixing with the KK tower of fermions, are realistic. 

\begin{figure}[t]
\center
\begin{center}
	\includegraphics[height=3.5cm]{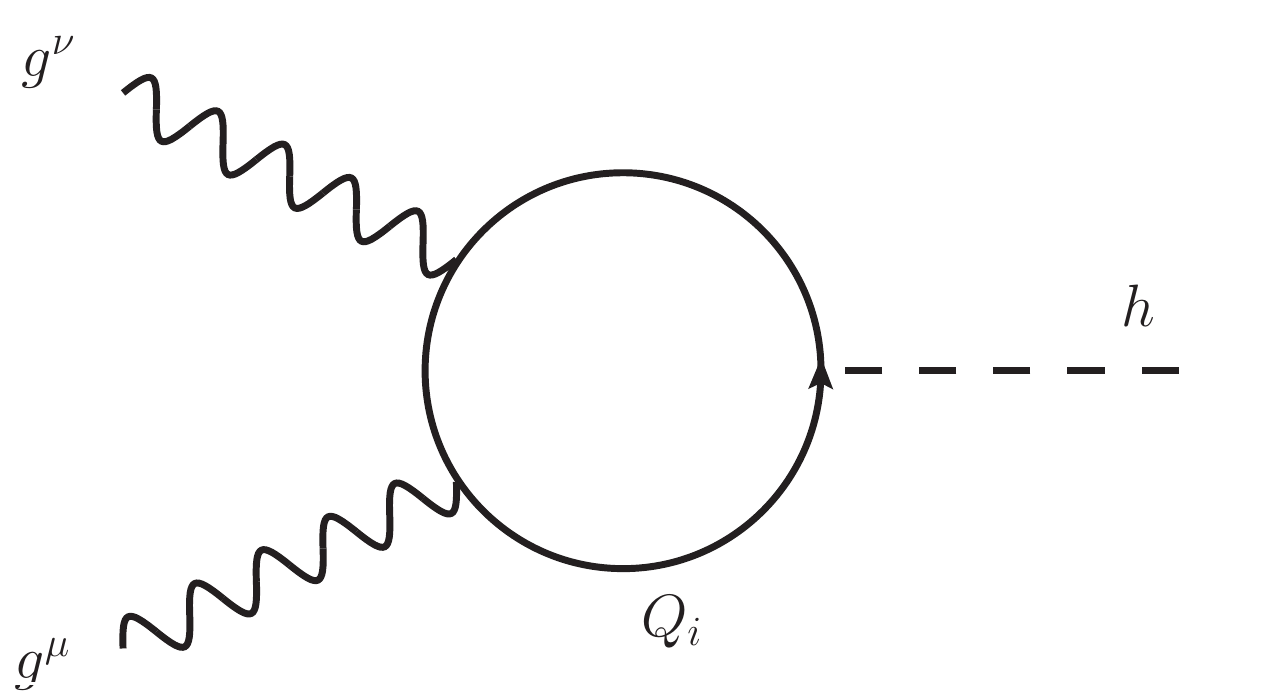}
 \end{center}
\vspace{.1cm}
\caption{Feynman diagram for the production cross section $gg \to h$ in warped space models.
 }  \label{fig:hgg}
\vspace{.4cm}
\end{figure}

The dominant contribution to the $hgg$ coupling is obtained at
one-loop level by calculating the diagram shown in Fig. \ref{fig:hgg}, with
fermions running in the loop.\footnote{We assume here for simplicity that the Yukawa
  couplings are real (no pseudoscalar component). The expressions
  for the general case can be found in Section V.} 
The diagram yields a cross section equal to \cite{Gunion:1989we}
\bea
\sigma_{gg\rightarrow h} = {\alpha_s^2 m_h^2\over 576 \pi} (c^S_{ggh})^2\ \delta(\hat{s} - m_h^2),
\ \ \ {\rm with}\ \ \ \ c^S_{ggh} = \sum_{f} {Y_f\over M_f} A^S_{1/2}(\tau_f), 
\eea
where $Y_f$ are the diagonal entries of the fermion Yukawa matrix in
the physical basis $ {\bf Y}^{phys}_u$, and $M_f$ are the diagonal
entries in  $ {\bf M}^{phys}_u$. In the case where we keep 2 full KK
levels ($N=2$) there are $30$ fermions in the loop (15 up-type and 15
down-type), while for $N=3$, there will be $42$ fermions in the loop,
including in both cases the 6 SM quarks.
Here   $\hat{s}$ is the invariant momentum squared of the gluons, $\tau_f\equiv
m_h^2/4m_f^2$  
and $A^S_{1/2}(\tau)$ is the spin-${1\over 2}$ form factor, given by
\bea\label{eq:loopfunction}
A^S_{1/2}(\tau) = {3\over2}\left[\tau + (\tau -1)f(\tau)\right]\tau^{-2}, \ \ \  
 f(\tau) = \left \{
  \begin{tabular}{cc}
  			 $\left[\arcsin\sqrt{\tau}\right]^2$ &  $(\tau\le 1)$\\
  			 $ -{1\over4}\left[\ln\left({1+ \sqrt{1-\tau^{-1}}\over 1- \sqrt{1-\tau^{-1}}}\right)-i\pi\right]^2$&  $(\tau> 1)$
\end{tabular}
\right.
\eea

\begin{figure}[t!]
\center
\begin{center}
\vspace{-2.5cm}
\hspace{-2cm}
	\includegraphics[width=18.0cm,height=9.5cm]{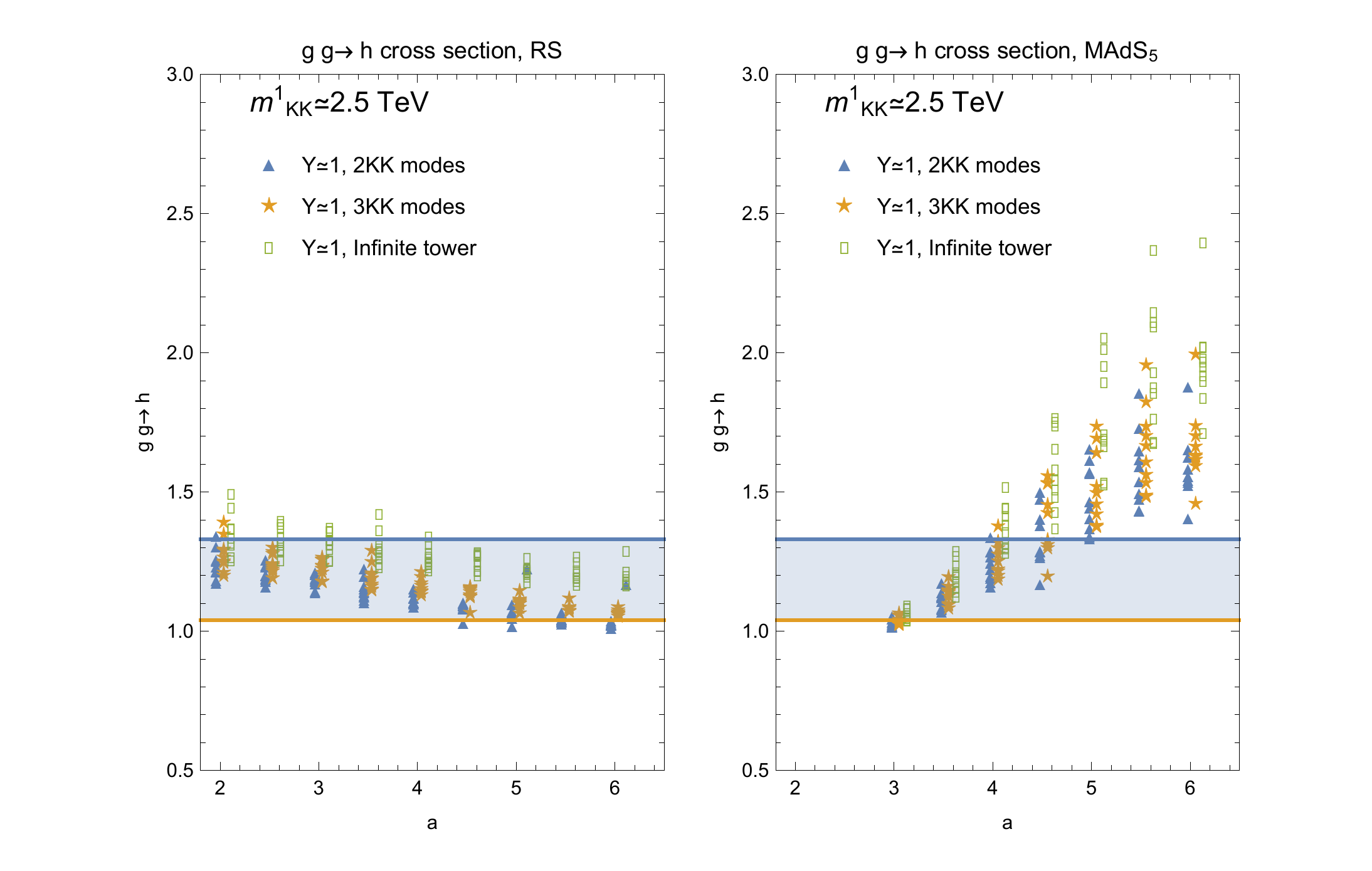} \\
\vspace{-.5cm}
\hspace{-2cm}
	\includegraphics[width=18.0cm,height=9.5cm]{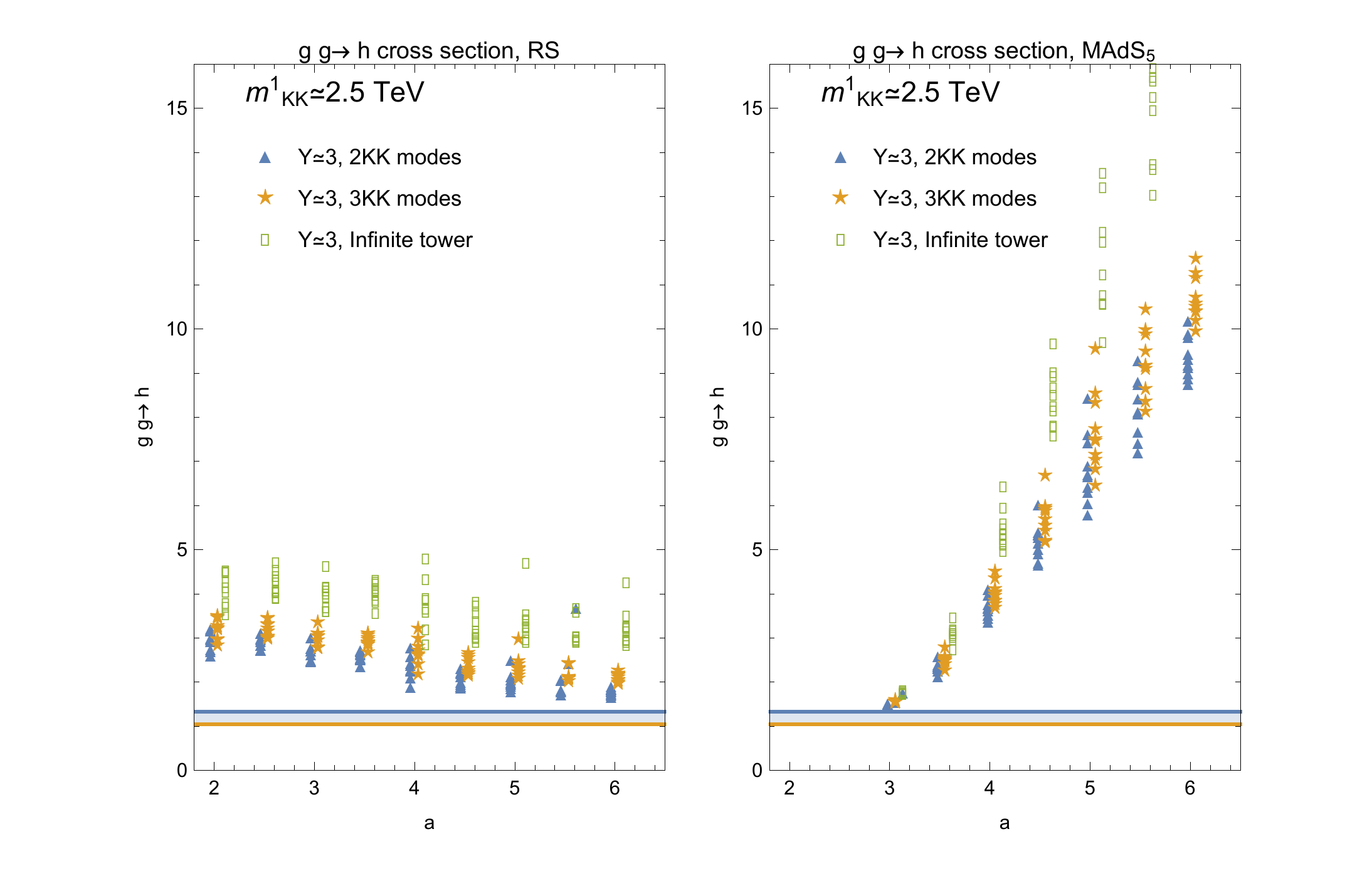}
 \end{center}
 \vspace{-1cm}
\caption{Higgs production rate ratio to Standard Model  prediction as
  a function of the Higgs localization parameter,  $a$. We consider an effective theory consisting of a
  tower of 2 (blue triangles) or 3 (orange stars) KK modes. We also include
  the results for an infinite tower of KK states (green squares). The
  KK masses are about $2.5$ TeV in both the RS metric (left panels)
  and  $MAdS_5$ scenario (right panels), for which we have
  chosen $\nu= 0.5$, $kL_1\simeq 0.2$. The 5D Yukawa couplings  are
  chosen such that  $Y^{5D}\sim1$ (upper panels) and $Y^{5D}\sim 3$
  (lower panels). The shaded regions show the experimental bounds from CMS and ATLAS. 
 }  \label{fig:hggCrossSection}
\end{figure}
In Fig. \ref{fig:hggCrossSection} we plot the Higgs production
cross section through gluon fusion. We show on the right panels, the full flavor
$gg\rightarrow h$ cross section in the modified  $AdS_5$ ($MAdS_5$)
metrics, and on the left side panels we show the results for the same
cross section for the model in the RS limit\footnote{To
  compare the two different metric scenarios, we keep
  $M_{\rm KK}=2.5$ TeV for both graphs. Note however that electroweak and precision
  tests force the KK scale to be much higher for models with RS
  metric.}. The top panels are for 5D Yukawa couplings such that
$Y^{5D}\sim 1$ and the bottom ones are for $Y^{5D}\sim3$.  Note that in order to
generate the top quark mass correctly, the $(Y_u^{5D})_{33}$ entry of
$Y_u^{5D}$ must always be of order $\sim 3$, even when the rest of
entries are taken to be $\sim 1$.


As the graphs indicate, the gluon fusion cross section is enhanced
compared to the SM one for both models, whether calculated with 2 or 3
 or infinite KK levels (see Sec. \ref{sec:4} for this last case). This confirms the expectation
from brane and bulk Higgs production with one flavor
\cite{Azatov:2010pf,Frank:2013un} and the results from summing over
the infinite tower with a bulk Higgs in \cite{Archer:2014jca}.


Note that in all panels of the graphs, the results obtained using a
finite number of KK levels and the results obtained with the complete
tower of KK levels start to deviate for larger values of $a$. This is
consistent with the findings in \cite{Frank:2013un} and it is linked
to an increased UV sensitivity of the scenario (i.e the decoupling of
heavy quark KK modes becomes harder and harder as the Higgs approaches
the IR brane). Also note that when the Yukawa couplings become larger, $Y^{5D} \sim 3$,
the predictions from the finite KK level calculation and from the
infinite KK levels also start to differ (signaling a potential issue). This result is again not 
surprising since our calculation of the infinite tower relies on the
smallness of a perturbative parameter $Y^2/M^2_{KK}$. As the Yukawa
couplings are taken larger and larger, the perturbative calculation becomes
worse, and this added to the effect of including the full flavor structure
which also increases the size of the perturbations, as 3 families of quarks are
now included for each KK level.
Even though this might seem as just a calculation problem, we must
be aware that we will be quickly approaching a non perturbative
limit of the theory, as the Yukawa couplings become strong (i.e. loops
including these couplings will start to become comparable to the tree level).
Note that in the RS limit, the  limit of low values the KK masses is in
conflict with electroweak precision bounds, so that for slightly
larger KK masses, one could allow for slightly higher Yukawa couplings.
In the  $MAdS_5$, strong coupling limit effects will become 
rapidly important as the value of the Higgs parameter $a$ grows, since
all the Higgs couplings grow exponentially in that limit.

The message from these investigations is therefore that a safe region of parameter
space (minimum UV sensitivity and safe from non-perturbative couplings) requires
moderate Yukawa couplings $Y^{5D}\sim 1$, as well as 
low Higgs localization parameter values, $a\sim 2-5$. As the graphs
show, in this region  both RS and $MAdS_5$ seem to be consistent
with LHC Higgs production data (except that in the RS limit scenarios are in
trouble with flavor and electroweak precision data for these low KK masses).

\begin{figure}[t]
\center
\vspace{-1.5cm}
\vspace{-1cm}
\hspace{-2cm}
	\includegraphics[width=18.0cm,height=9.5cm]{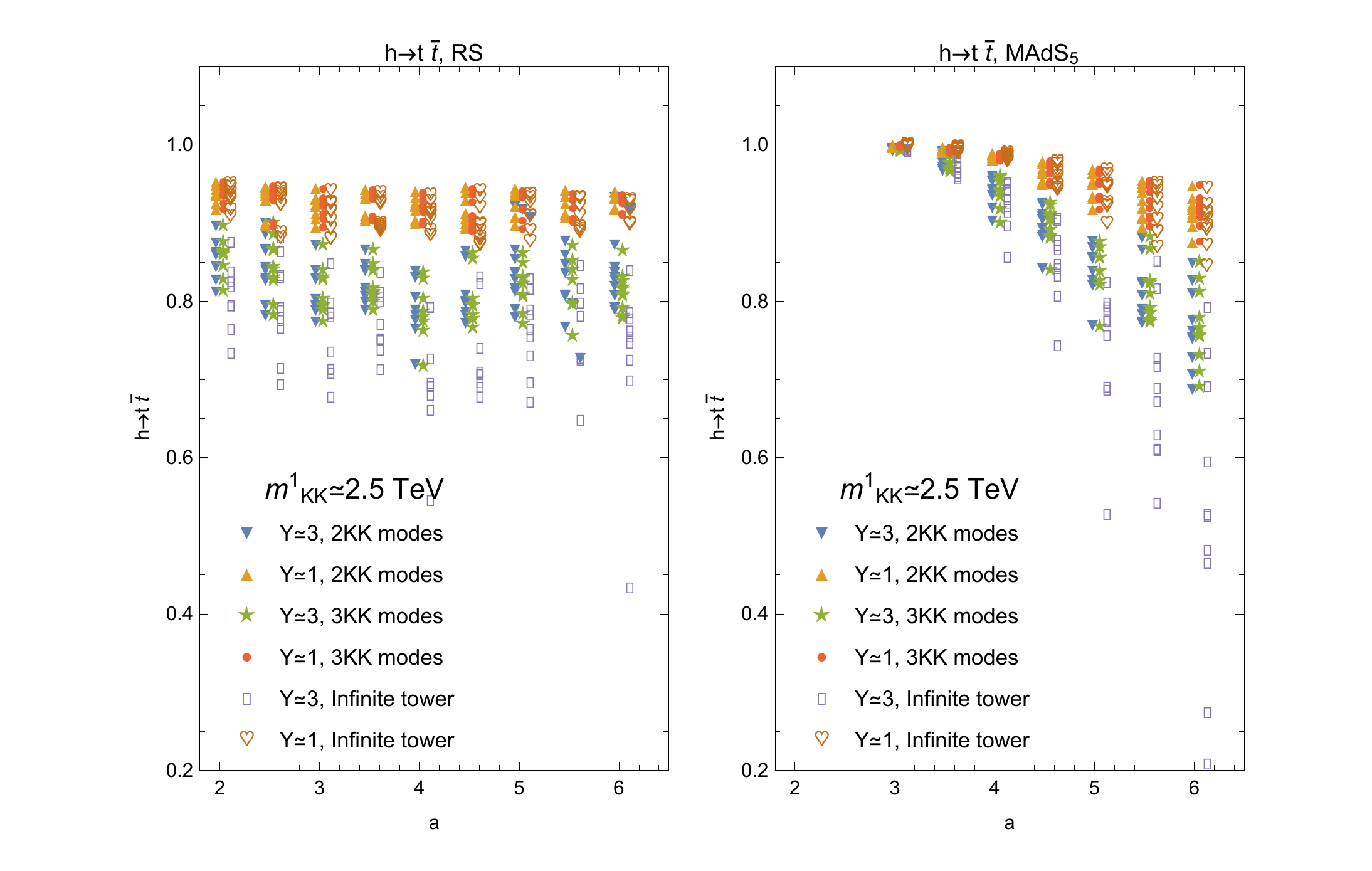}\\
\vspace{-.5cm}
\caption{Top quark Yukawa couplings relative to their SM values as a function of the Higgs
  localization parameter, $a$, in  the  $MAdS_5$ scenario (right), and its RS limit (left). We have considered an
  effective field theory consisting of a tower of 2 (triangles) and 3 (stars and dots) KK 
levels with $Y^{5D}\sim 1$ (warm colors) and $Y^{5D}\sim 3$ (cold colors), 
with the lightest KK mass at about $2.5$  TeV. For both graphs we include also an infinite tower of KK states (hollow shapes).
 For the general metric scenario, we have chosen $\nu= 0.5$,
  $kL_1\simeq 0.2$.      } 
\label{fig:topyukawaShift}
\end{figure}

\begin{figure}[t!]
\center
\vspace{-1.5cm}
\hspace{-2cm}
	\includegraphics[width=18.0cm,height=9.5cm]{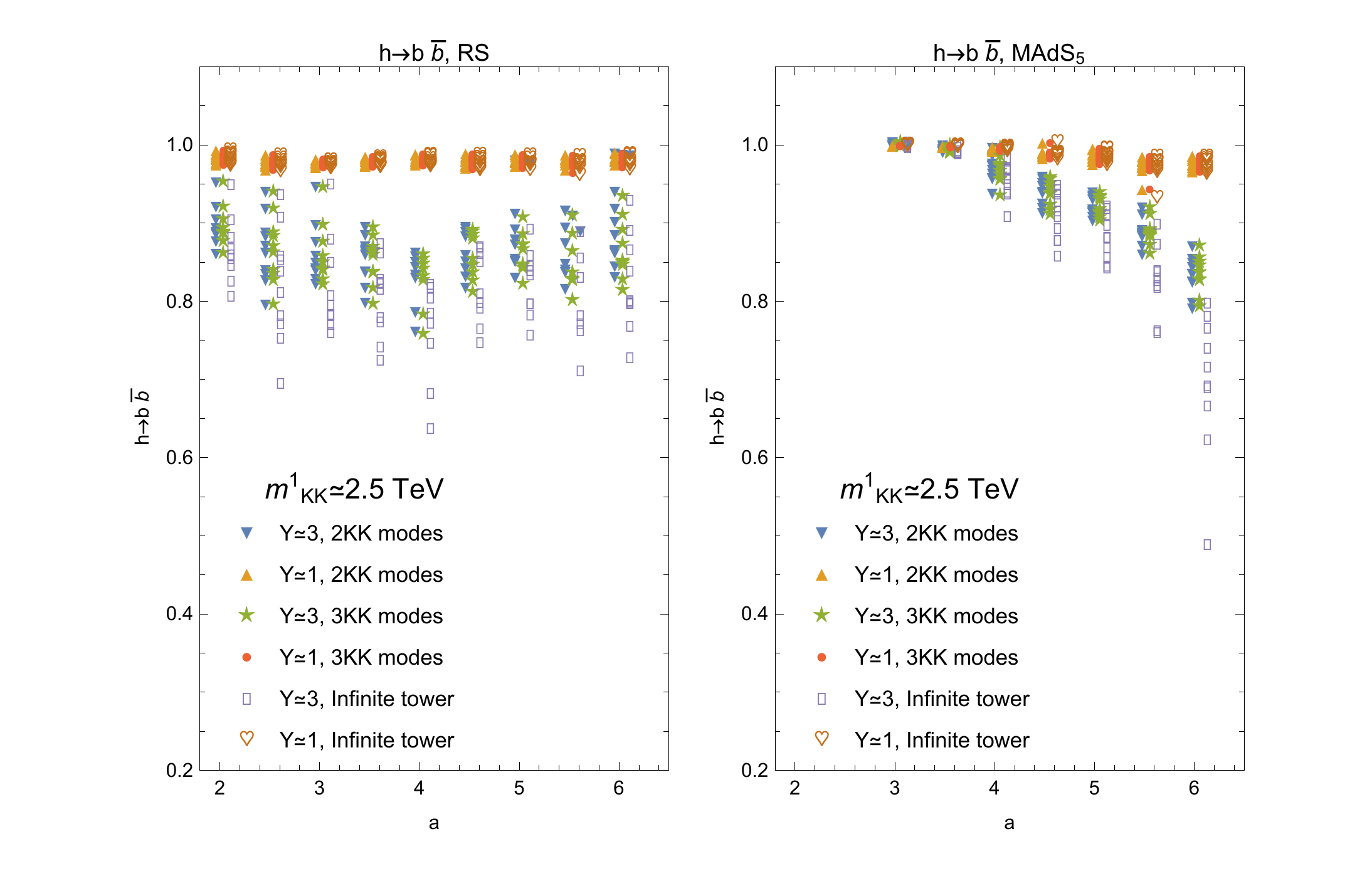}\\
\vspace{-.7cm}
\hspace{-2cm}
	\includegraphics[width=18.0cm,height=9.5cm]{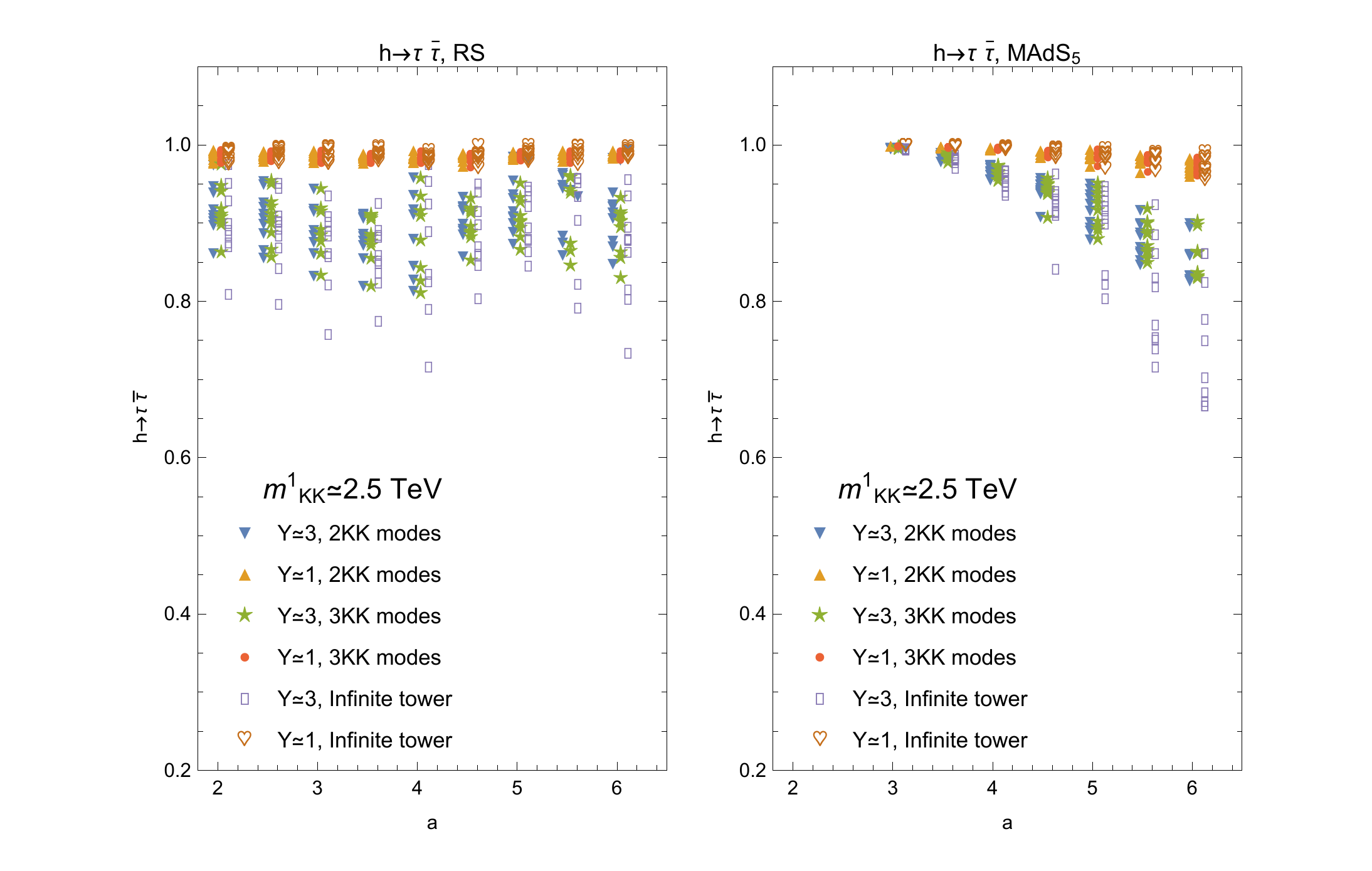}
\caption{Bottom quark Yukawa couplings 
  (upper panels) and tau lepton Yukawa couplings (lower panels),  relative to their SM values, as functions of the Higgs
  localization parameter, $a$, in  the  $MAdS_5$ scenario (right), and its RS limit (left). We considered a tower of 2 (triangles), 
  3 (dots and stars) KK modes and an infinite  KK tower (hollow shapes).} 
\label{fig:yukawaShift}
\end{figure}

In Figs. \ref{fig:topyukawaShift} and \ref{fig:yukawaShift} we have plotted the deviation of the
physical top Yukawa coupling with respect to its SM value, and the same for 
the bottom quark and the tau lepton Yukawa couplings. Again the 
$MAdS_5$ metric scenario is shown on the right side and the RS limit on the left. The
consistency between the results of the different calculations (2 KK vs. 3 KK vs. full
KK tower) is better here than for the loop-dominated graphs in which
all families of KK quarks contributed evenly to the results
(essentially as a trace),  whereas for single SM quark coupling
essentially only one KK tower contributes.

Although their effect to the gluon fusion cross
section is not as important as that of the top quark, measurements of
the $hb\bar b$ and $h\tau^+ \tau^-$ are underway, and higher precision
at the Run II of the LHC means that these could be compared to the
experimental data in the near future.  

\begin{figure}[t]
\center
%
\vspace{-2cm}
\hspace{-2cm}
	\includegraphics[width=18.0cm,height=9.5cm]{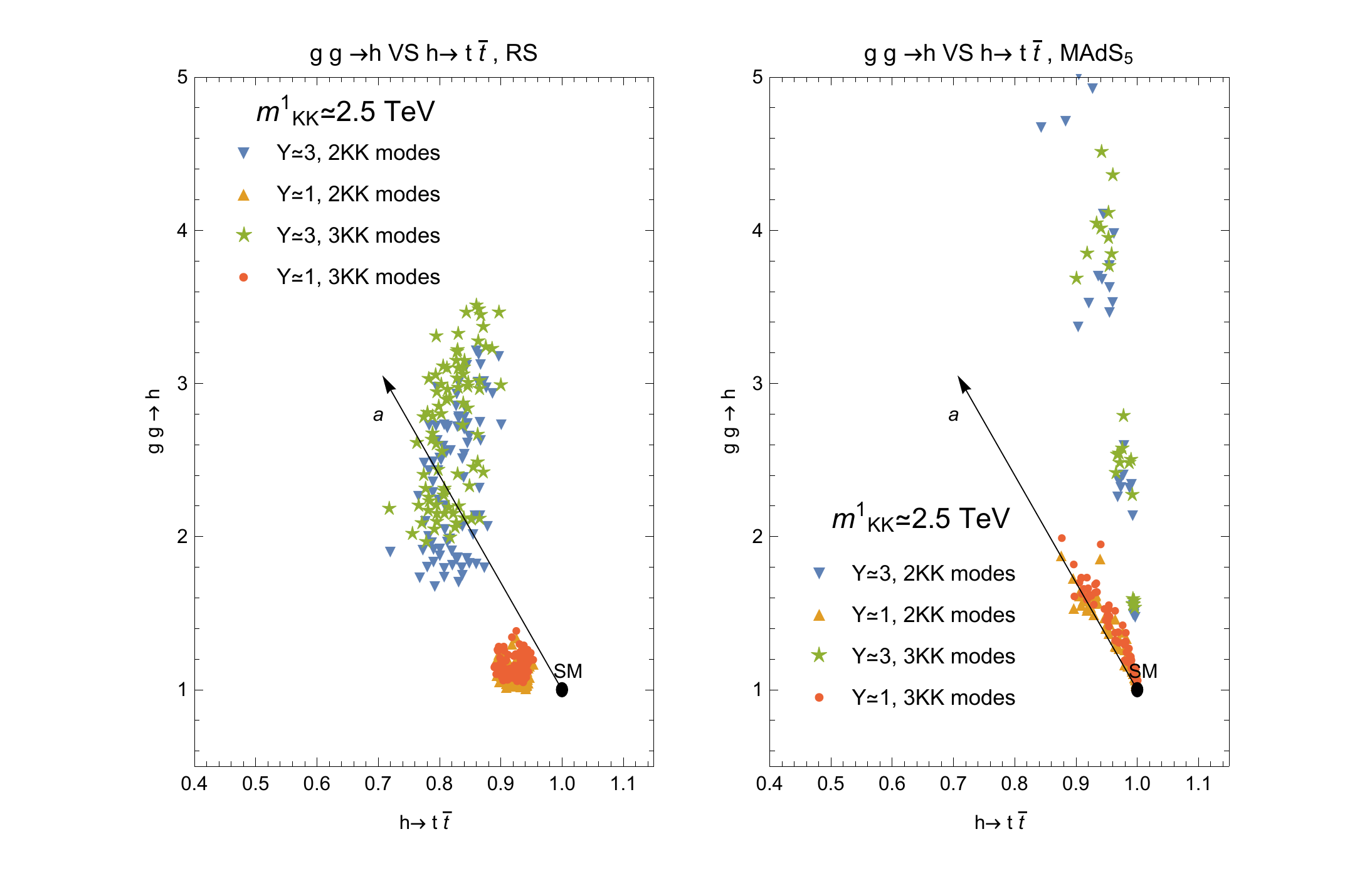} 
\caption{Comparison of the production cross section $gg \to h$ versus the Yukawa couplings for the top quark in $MAdS_5$ (right) and the corresponding RS limit (left) a tower of 2 (triangles), 3 (dots and stars) KK modes. The value of the $a$ parameter decreases in both plots, from left to right along the diagonal.} 
\label{fig:yukawaShifts_vs_gg}
\end{figure}

Finally in Fig. \ref{fig:yukawaShifts_vs_gg} we show a comparison between
the top Yukawa couplings and the gluon fusion cross section in
$MAdS_5$ (right) and the corresponding RS limit (left). In the figure,
the values for $a$ parameter decrease from left to right along the
diagonal in both plots. The SM values, shown as black blobs are
indicted in the bottom right-hand corner of the plots and appear to
overlap with most of the $Y \simeq 1$ parameter points, more so for the $MAdS_5$
scenario. The enhancements compared to the SM come mostly from the KK
loop-enhanced gluon fusion, as explained above. 


For the plots in Fig. \ref{fig:yukawaShifts_vs_gg} we have set the 5D Yukawa couplings, $Y^{5D} \sim 1$, and considered two models, 
the  $MAdS_5$ model  and its pure $AdS_5$ limit, \ie  ~RS. 
Our generic metric model has metric parameters such that $\nu = 0.5$, $ky_1 = 0.2$, 
and KK masses $m_{KK} \sim 2.5$ TeV. For the RS limit, we have taken $\nu = 10$, $ky_1 = 0.9999$
and $m_{KK} \sim 2.5$ TeV. 

Our results show that the  $MAdS_5$ models are more sensitive
to the values of the $a$-parameter and, while for small values of the
$a$-parameter, \ie ~a delocalized Higgs field, the model is in
accordance  with the current experimental bounds on the Higgs
production rate through gluon fusion,  for larger values of $a$ the
enhancement increases. The shift in the Yukawa couplings also
exhibit the same $a$-parameter dependence. This result is consistent
with our previous results for one generation
\cite{Frank:2013qma}. Note that the Yukawa couplings for both cases of a
top-like fermion and an up-like fermion are suppressed, and the reason
for the observed enhancement in the $gg\rightarrow h$ cross section is
due to the running of KK modes in the loop of the Feynman diagram in Fig. \ref{fig:hgg}
\cite{Azatov:2010pf}. 

Intuitively, the reason for this dependence on the localization of the Higgs field in the 
$MAdS_5$ models is due to the fact that in these models the volume of
the fifth dimension is smaller, as shown in Fig. \ref{fig:TA}. 
The deviation from the $AdS_5$ near the IR brane results in a more aggressive warping 
of space near that brane. As a consequence all the KK modes, including KK fermions, are 
pushed more towards the IR brane, which results in smaller values of the overlap integrals 
of Eq. (\ref{YYY1KK}) for a delocalized Higgs field. On the other hand, for the same reason, 
as the Higgs field becomes more and more localized on the IR brane,  deviations become 
more substantial for the  $MAdS_5$ scenario compared to the RS-like limit.

\section{$h \to \gamma\gamma$ decay}
\label{sec:3}
\begin{figure}[t]
\center
\begin{center}
\hspace{-0.5cm}	\includegraphics[height=3cm]{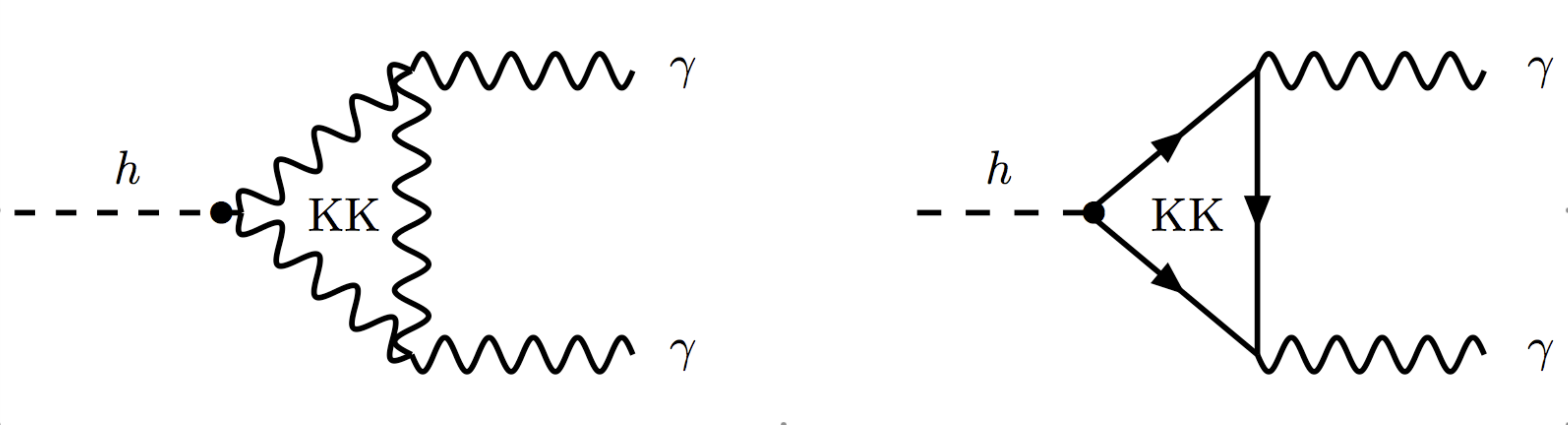}
 \end{center}
\caption{Feynman diagrams for $h\rightarrow \gamma\gamma$ in $MAdS_5$. } 
\vspace{.4cm}
\label{fig:hgaga}
\end{figure}

The diagrams responsible for the decay $h \to \gamma \gamma$ in $MAdS_5$ are shown in Fig. \ref{fig:hgaga}. 

The decay width from the diagrams, 
where the KK partners to $W^{\pm}$ bosons run in loop of the left side  diagram, and the KK fermions  in the right side diagram,
respectively, is given by
\cite{HiggsReview, Bouchart:2009vq}
\bea\label{hgammagamma}
\Gamma_{h \to \gamma\gamma} = {{\alpha^2 m_h^3}\over{256 \pi^3}}{1\over{v^2}} 
\left| \sum_{n} {{g^{\pm}_{W^n}}\over{m_{W^n}^2}} A^h_{1}(\tau_{W^n}) + 
{4\over3}\sum_{\{f\}} {{Y_f}\over{M_f}} N_c Q_f^2 A^h_{1/2}(\tau_f) \right|^2, 
\eea
where $n = \{0,1,2,\dots\}$ for the zero and the corresponding KK modes and $f$ runs over all the fermions 
and their corresponding KK partners, $N_c = 3$ for quarks and 1 for leptons, respectively, and $Q_f$ 
is the charge of the fermion in the loop. The form factor $A_{1/2}$ is previously presented in Eq. \ref{eq:loopfunction} 
and for spin-1 bosons in the loop, $A_1$, is given by
\bea
A^h_{1}(\tau) = -[2\tau^2 + 3\tau + 3(2\tau-1)f(\tau)]\tau^{-2}.
\eea
The W boson zero mode and KK masses, $m_{W^n}^2$ are given by the diagonalization of the
mass terms, ${\cal M}^2_\pm$ in the Lagrangian
\bea\nonumber
{\cal L}_{\rm mass}^{\rm c} \! = \! (W^+_\mu, W^{+(1)}_\mu, W^{+(2)}_\mu  \ \ \dots) \ {\cal M}^2_\pm  \
( W^{-\mu}, W^{-(1)\mu}, W^{-(2)\mu}  \dots )^T,
\eea
where the mass matrix and all the gauge boson fields are in the gauge basis. Keeping only the first
two KK modes for  numerical calculations,  we have, in the mass basis
\bea\label{massDiag}
\mathcal{M'}^2_\pm \equiv V \ {\cal M}^2_\pm \ V^\dagger = {\rm diag}~(m^2_W,m^2_{W^1},m^2_{W^2}) .
\eea
Using the  transformation
as in Eq. (\ref{massDiag}) to the coupling matrix of the gauge bosons in the gauge basis,
\bea
{\cal L}_{\rm coupling}^{\rm c} \! = 2 h (W^+_\mu, W^{+(1)}_\mu, W^{+(2)}_\mu\ \dots \ ) \ {\cal G}_\pm  \
( W^{-\mu}, W^{-(1)\mu}, W^{-(2)\mu} \ \ \dots)^T ,
\eea
we obtain 
\bea  
{\cal L'}_{\rm coupling}^{\rm c} \  = \ 2 \ h \ (W_{\mu} \ \ W^1_{\mu}  \ W^2_{\mu} \ \ \dots ) \ {\cal G'}_\pm \
(W^{\mu} \  W^{1\mu} \ \ W^{2\mu} \ \ \dots ) ^T,   
\eea
with 
\bea
{\cal G'}_\pm \ = \ V \ {\cal G}_\pm \ V^\dagger.
\eea
The gauge couplings, $g^{\pm}_{W^n}$ in Eq. (\ref{hgammagamma}) are then given by the diagonal elements of 
the matrix ${\cal G'}_\pm$. 
\begin{figure}[t]
\center
\begin{center}
\vspace{-2.3cm}
\hspace{-2cm}
	\includegraphics[width=16.0cm,height=8.5cm]{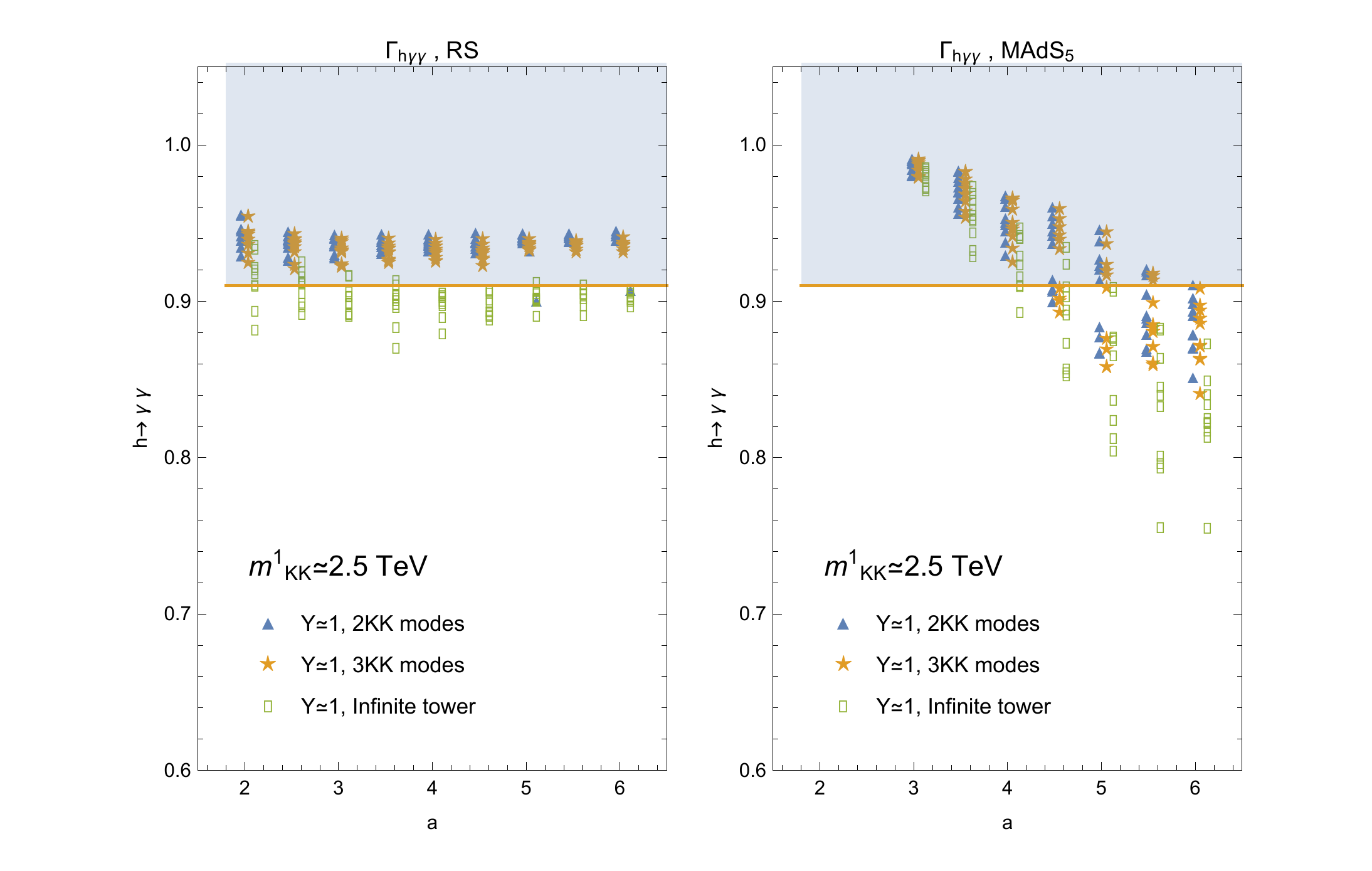}\\
\vspace{-.3cm}
\hspace{-2cm}
	\includegraphics[width=16.0cm,height=8.5cm]{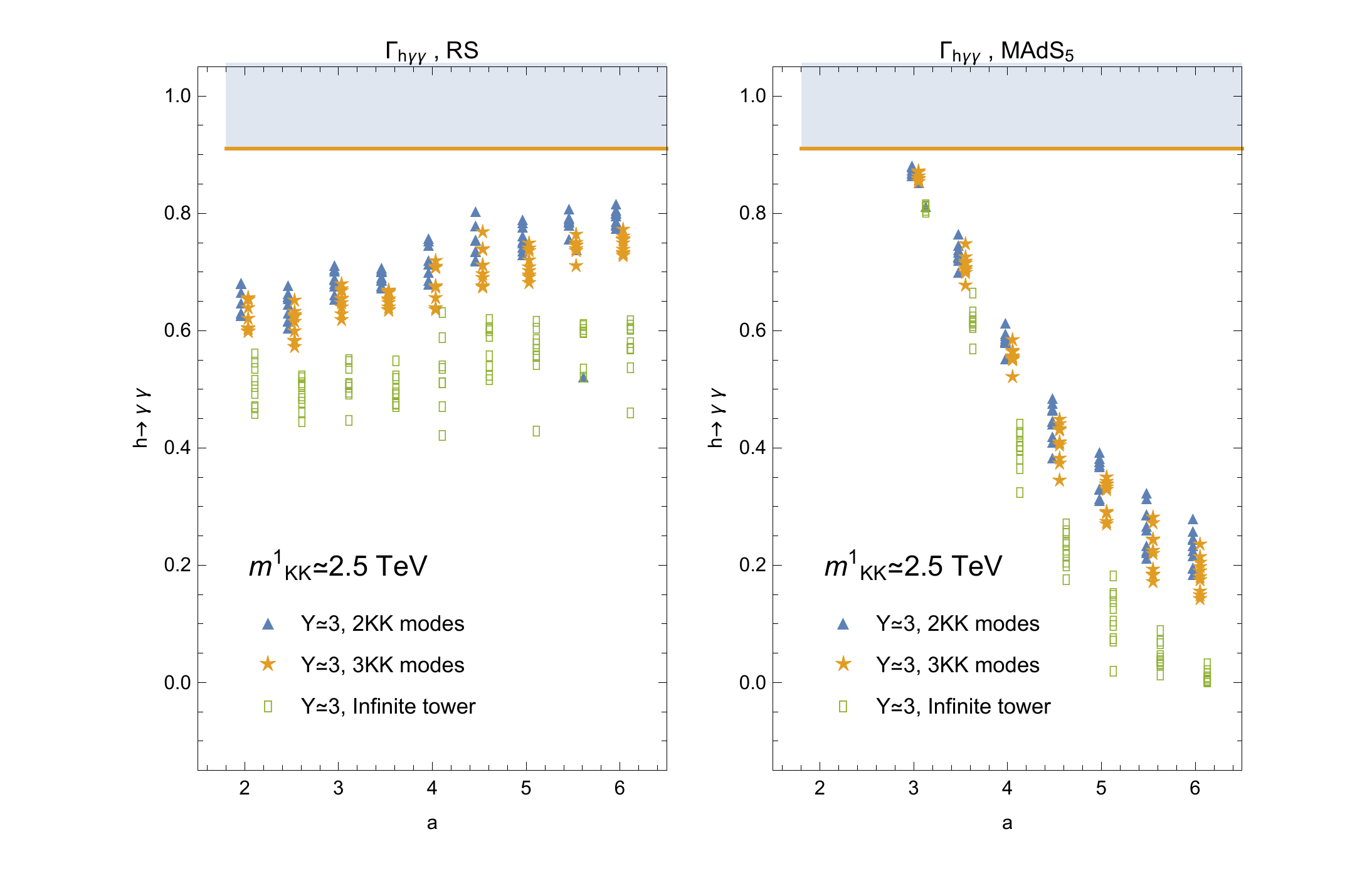}
 \end{center}
 	\vspace{-1.cm}
\caption{Plot of the decay width for $h \to \gamma \gamma$ in $MAdS_5$
  relative to SM width, as a function of the Higgs
  localization parameter,   $a$. In all scenarios we consider an
  effective field theory consisting of a   tower of 2  (triangles) or
  3 (stars) KK modes.  We also include the calculation in which
  an infinite tower of KK states in included (hollow squares). For the
  $MAdS_5$ (right panels)  we have chosen
  $\nu= 0.5$, $kL_1\simeq 0.2$. The 5D Yukawa
  couplings are chosen such that $Y^{5D}\sim1$ (top panels) and $Y^{5D}\sim 3$ (bottom panels). The left-hand side
  shows, for comparison, the same values for the RS  limit of the model. The lightest KK  mass is   about
  $2.5$ TeV in both scenarios. }  
\label{fig:Hgammgamm}	
\vspace{-0.1cm}
\end{figure}

In Fig. \ref{fig:Hgammgamm} we show the decay width for $h \to \gamma \gamma$ 
relative to the one in the SM,  as a function of the Higgs
localization parameter, $a$, for the effective theory containing a
tower of 2 (triangles) or 3 (stars) KK modes. We include here, as
well, results of the calculation with the infinite KK tower, for the
same masses, as explained in the next section, Sec. \ref{sec:4}, as
hollow squares. We plot the values for the $MAdS_5$ scenario
(right panel) and, for comparison, the same values for the
RS limit of the model (left panel).    The parameters have been chosen as
$\nu= 0.5$, $kL_1\simeq 0.2$   and the lightest KK mass is   about
$2.5$ TeV, in both RS and $MAdS_5$ scenarios. The 5D Yukawa
couplings   are chosen such that $Y^{5D}\sim1$ for the top panels and
$Y^{5D}\sim3$ for the bottom panels. Both models show a suppression of
the di-photon decay widths with respect to the SM values, consistent with the one-generation results
for RS with fields on the brane in \cite{Azatov:2010pf}, and for the
three-generation in RS with the Higgs field in the bulk in
\cite{Archer:2014jca}. Again, the convergence is better for low $a$ values
 than for a Higgs localized more towards the brane (larger $a$ values). Note again the discrepancy of
the infinite tower calculation in RS limit for $Y=3$ noted in the
previous section and linked to approaching a non-perturbative limit
for these low KK masses. 
 
\begin{figure}[t]
\center
\begin{center}
\vspace{-1.5cm}
\hspace{-2cm}
	\includegraphics[width=18.0cm,height=10cm]{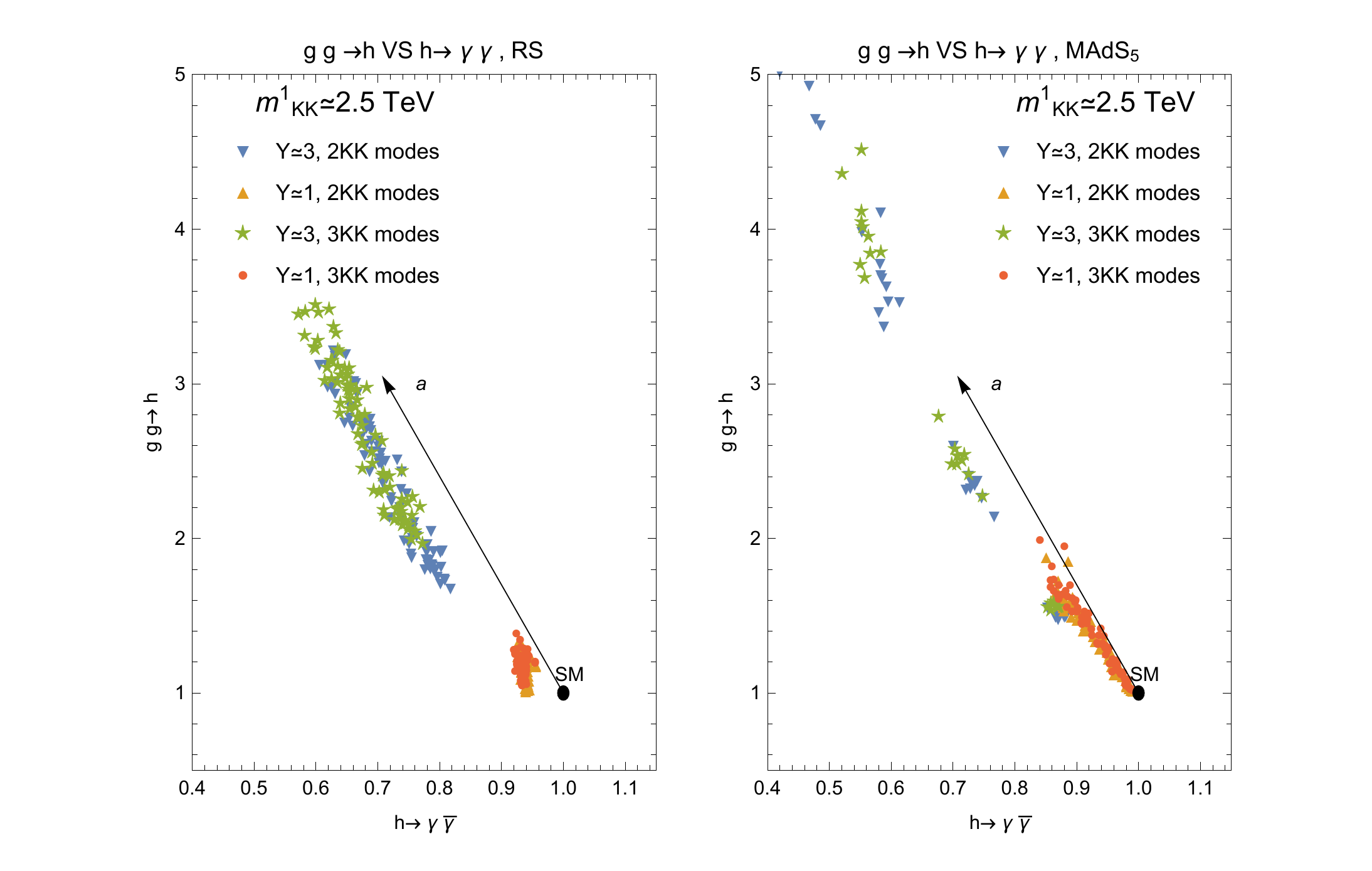}
 \end{center}
	\vspace{-.4cm}
\caption{Comparison of the production rate through gluon fusion $gg
  \to h$ versus the di-photon decay rate in $MAdS_5$ for an effective
  field theory consisting of a tower of 2 (triangles) or 3 (dots and stars) KK
  modes. The parameters are chosen as in Fig. \ref{fig:Hgammgamm}. The
  SM values are shown for comparison as a black dot. The right panel
  shows the $MAdS_5$ model, compared to the left hand side as RS limit
  of the model.} 	\vspace{-.4cm} 
\label{fig:Hgg_vs_Hgammgamm}
\end{figure}

 Fig. \ref{fig:Hgg_vs_Hgammgamm} shows the comparison between the
 loop-dominated production $gg \to h$ and decay $h \to \gamma \gamma$
 for the $MAdS_5$ model (right panel) and its RS limit (left
 panel). The correlation function is almost linear for RS. Again the
 localization parameter $a$ decreases along the diagonal, from left to
 right and convergence is better for small $a$'s. The $MAdS_5$
 parameter points overlap with the SM values (solid black circle) for
 $Y \simeq 1$ (orange region) and even for $Y \simeq 3$, low $a$, while
 the RS limit has a much smaller region of approximate agreement, and
 that is only true for $Y \simeq 1$.

\section{Infinite tower with full flavor contribution}
\label{sec:4}

In this section we tackle the calculation of the Higgs couplings when considering 
the complete towers of KK fields along with the full
flavor structure of the SM. The purpose of this exercise is 
to obtain an independent result from the previous formalism
(with two full-flavored KK levels and three full-flavored KK levels) as well as a good
check on the decoupling of heavy degrees of freedom in different parts
of the parameter space. It is also important to note that the full tower
calculation presented here is performed perturbatively (in terms of
$Y^2 v^2 / M_{KK}^2$), so that for larger Yukawa couplings the
convergence of the expansion should worsen.  
We set up the calculation for the up-quark sector, with the
understanding that it can be trivially extended for both the down-quark
sector and the charged lepton sector. 

We introduce a set of three
families of 5D $SU(2)$ quark doublets $Q^i(x,y)=q_L^i(x,y)+q_R^i(x,y)$
with 5D family index $i=1,2,3$, as well as three quark singlets
$U^i(x,y)=u_R^i(x,y)+u_L^i(x,y)$, where $x$ represents 
the 4D spacetime variables and $y$ the extra dimension. 
 We perform the dimensional reduction as usual, by setting the
following separation of variables for the different 5D quark fields:\\
$\ \ \displaystyle q^i_L(x,y) = \sum_{n=0}^{3N} q_L^{i n}(y)\, q_L^n(x)$,
$\ \displaystyle q^i_R(x,y)=\sum_{n=0}^{3N}q_R^{in}(y)u^n_R(x)$,
$\  \displaystyle u^i_L(x,y)  = \sum_{n=0}^{3N}  u_L^{in}(y)  q_L^n(x)$, and
$\  \displaystyle  u^i_R(x,y) =\sum_{n=0}^{3N}  u^{in}_R (y)\, u_R^n
(x)$.

Note that the $n=0$ states ($q_L^0 (x)$ and $ u^0_R(x)$) are the SM doublet and singlet 
up-quark, and contain a piece of each of the three bulk families,
represented by the twelve wavefunctions $q_L^{1 0}(y)$, $q_L^{2
  0}(y)$, $q_L^{3 0}(y)$, $q_R^{1 0}(y)$ etc$\dots$ . Successively, $n=1$
corresponds to the SM charm quark and $n=3$ to the SM top
quark, each of them carrying a mixture of twelve wavefunctions. Higher
values of $n$ correspond to heavy KK quarks. Within the warped metric
background of Eq.~(\ref{metricansatz}), the coupled equations 
of motion corresponding to the 12 wavefunctions for each KK level $n$ are
\bea
m_n^*q_L^{in}+e^{A+Q_{q_i}}\partial_y(q_R^{in}
e^{-2A-Q_{q_i}})-e^{-A}v(y)\sum_{j=1}^{3}\frac{Y^*_{ij}}{\sqrt{k}}u^{jn}_R&=&0 \label{eq1}\\
m_nq_R^{in}-e^{A-Q_{q_i}}\partial_y(q_L^{in} e^{-2A+Q_{q_i}})-e^{-A}v(y)\sum_{j=1}^{3}\frac{Y^*_{ij}}{\sqrt{k}}u^{jn}_L&=&0 \label{eq2} \\
m_n^*u_L^{in}+e^{A+Q_{u_i}}\partial_y(u_R^{in}
e^{-2A-Q_{u_i}})-e^{-A}v(y)\sum_{j=1}^{3} \frac{Y_{ji}}{\sqrt{k}} q^{jn}_R&=&0 \label{eq3}\\
m_nu_R^{in}-e^{A-Q_{u_i}}\partial_y(u_L^{in}
e^{-2A+Q_{u_i}})-e^{-A}v(y)\sum_{j=1}^{3} \frac{Y_{ji}}{\sqrt{k}} q^{jn}_L&=&0 \label{eq4}
\eea
where we defined
\bea
Q_i(y) \equiv \int M_i(y) dy,\non
\eea
with $M_i(y)$ being the 5D bulk mass associated to each 5D fermion. For
simplicity and to maintain an analogy with the usual bulk RS scenario, we take
\bea
Q_i(y) \equiv c_i A(y)
\eea
with $c_i$ being the bulk fermion $c$-parameters similar to the RS ones.

From the previous equations we can obtain the following exact
relations for the effective 4D mass $m_n$ of the $n$-th KK level  as well
as its effective 4D diagonal Yukawa coupling $y_{nn}$
\bea
 m_{n} &=& \sum_{i,j=1}^3 \int dy \Big( e^{-3A} m_{n}  \left(u^{in*}_L u_L^{in} +
 q_R^{in*}q_R^{in} \right) +  e^{-4A} {v(y)} \left(q_L^{ in} \frac{Y_{ij}}{\sqrt{k}}
 u_R^{j n*} - q_R^{in*} \frac{Y^*_{ij}}{\sqrt{k}} u_L^{ j n}  \right) \Big)    \hspace{1cm}\label{mn}\\ 
 y_{nn} &=& \sum_{i,j=1}^3 \int dy  e^{-4A} {h(y)} \left(q_L^{ in} \frac{Y_{ij}}{\sqrt{k}}
 u_R^{j n*} + u_L^{in} \frac{Y^*_{ji}}{\sqrt{k}} q_R^{ j n*}  \right) \Big)    \hspace{1cm}\ \label{yukawann}
\eea
where $v(y)$ and $h(y)$ are the profiles of the Higgs VEV and of the
lowest Higgs KK state (i.e the profile of the SM Higgs field).
The shift between the mass term and the diagonal Yukawa term is
\bea
\Delta_{n}\ =\ m_n-v_4 y_{nn}
\ =\  \sum_{i,j=1}^3 \int dy   \Big( e^{-3A}  m_{n}  \left(u^{in*}_L u_L^{in} +
 q_R^{in*}q_R^{in} \right) - 2   e^{-4A} {v(y)} q_R^{in*} \frac{Y^*_{ij}}{\sqrt{k}} u_L^{ j n} \Big)  
\eea
The off-diagonal 4D effective Yukawa coupling between the Higgs boson and
two fermions of levels $n$ and $m$ is 
\bea
 y_{nm} = \sum_{i,j=1}^3 \int dy  e^{-4A}{h(y)} \left(q_L^{ in} \frac{Y_{ij}}{\sqrt{k}}
 u_R^{j m*} + u_L^{in*} \frac{Y^*_{ji}}{\sqrt{k}} q_R^{ j m*}  \right) \Big)    \hspace{1cm}\ \label{yukawanm}
\eea
We can thus calculate these terms if we know the solutions for the
profiles from the coupled equations (\ref{eq1}), (\ref{eq2}), (\ref{eq3}) and
(\ref{eq4}). 

The strategy we follow is to solve them perturbatively for the case of
light (SM) modes, and therefore obtain masses and Yukawa couplings for the
SM up quark, charm quark and top quark, i.e. the levels $n=0,1,2$. We obtain 
\bea
q_L^{in}(y)&\simeq&e^{2A-Q_{q_i}}\left(Q_L^{in}+ m_n \int  q^{in}_R(y)
e^{-A+Q_{q_i}}  - \sum_j \frac{Y^*_{ij}}{\sqrt{k}} \int e^{-2A+Q_{q_i}} v(y) u^{jn}_L(y) \right)\label{ql}\\
u_R^{in}(y)&\simeq&e^{2A+Q_{u_i}} \left(U_R^{in}-m_n^* \int  u^{in}_L(y)
e^{-A-Q_{u_i}}  + \sum_j \frac{Y_{ji}}{\sqrt{k}} \int e^{-2A-Q_{u_i}} v(y) q^{jn}_R(y)
\right) \label{ur}\\
q_R^{in}(y)&\simeq&e^{2A+Q_{q_i}}\left(-Q_L^{in}m_n^*\int_0^ye^{A-2Q_{q_i}}+\sum_j
\frac{Y_{ij}^*}{\sqrt{k}}U_R^{jn}\int_0^ye^{Q_{u_j}-Q_{q_i}}v(y)\right)\label{qr}\\ 
u_L^{in}(y)&\simeq&e^{2A-Q_{u_i}}\left(U_R^{in}m_n\int_0^ye^{A+2Q_{u_i}}-\sum_j
\frac{Y_{ji}}{\sqrt{k}}Q_L^{jn}\int_0^y e^{Q_{u_j}-Q_{q_i}}v(y)\right), \label{ul}
\eea
where the 6 constants of integration $Q_L^{in}$ and $U_R^{in}$ (with $i=1,2,3$ and with the level $n$ fixed) are obtained after imposing Dirichlet boundary conditions on the
appropriate wave functions.


We are now equipped to compute the couplings of the Higgs boson 
with the SM fermions, including this time the effects of the full tower of KK
modes. Using these we can address the Higgs production calculation
with full KK towers, with the caveat that it is a perturbative calculation, whose
convergence should worsen for larger values of the 5D Yukawa couplings.


The radiative couplings of Higgs to gluons will depend on the physical Yukawa
couplings $y_{nn}$ of all the fermions running in the loop and on their physical
masses $m_n$. The real and imaginary parts of the couplings (scalar
and pseudoscalar parts) are associated with different loop functions, $A^S_{1/2}$ and $A^P_{1/2}$, as they
generate the two operators $h G_{\mu\nu}G^{\mu\nu}$ and $h
G_{\mu\nu}\tilde{G}^{\mu\nu}$. 

The cross section is 
\bea
\sigma_{gg\rightarrow h} = {\alpha_s^2 m_h^2\over 576 \pi} \left[|c^S_{ggh}|^2 + |c^P_{ggh}|^2\right]\ 
\eea
where
\bea
 c^S_{ggh} =\sum_n \re\left({\frac{y_{nn}}{m_{n}}}\right) A^S_{1/2}(\tau_f)
\hspace{.6cm} {\rm and} \hspace{.6cm}
c^P_{ggh} = \sum_n \im\left(\frac{y_{nn}}{m_{n}}\right) A^P_{1/2}(\tau_f)
\eea
with $\ \tau = m^2_h/4m^2_n\ $, with $A^H_{1/2}(\tau)$ defined as in
Eq.(\ref{eq:loopfunction}) and with
$\displaystyle\ \  A^H_{1/2}(\tau)=-f(\tau)\tau^{-2}$ \cite{Gunion:1989we}.
%
%

For heavy KK quarks with masses $m_n$ much greater than the Higgs mass
$m_h$ (i.e. when $\tau$ is very small) the loop functions are essentially
constant, as they behave asymptotically as $\displaystyle \lim_{\tau
  \to 0} A^S_{1/2} =1 \hspace{.2cm} \mbox{and} \hspace{.2cm}
\lim_{\tau \to 0} A^P_{1/2} =3/2.$ On the other hand, for light quarks (all the SM quarks
except top and bottom), the loop functions essentially vanish 
asymptotically  as $\displaystyle \lim_{\tau \to \infty} A^S_{1/2}
= \lim_{\tau \to \infty} A^P_{1/2} = 0.$  

In those two limits, the amplitudes $c^S_{ggh}$ and $c^P_{ggh}$ can be
written in terms of traces involving the infinite fermion mass and Yukawa matrices of the up and down
quark sectors, ${\bf M}_i$ and ${\bf Y}_i$ with $i=u,d$. 
Since the trace is basis invariant, we consider the infinite mass
matrices in the gauge basis ${\bf M}_u$ and ${\bf M}_d$ as defined in
Eq.~(\ref{mumatrix}), but this time for the case $N\to \infty$.

The infinite up-type Yukawa matrix can be obtained as
$\displaystyle {\bf Y}_u = \frac{\partial {\bf M}_u}{\partial v}$, and
so the traces we want to evaluate can be written as
\bea
\sum_n \re\left({\frac{y^u_{nn}}{m^u_{n}}}\right)
=\mbox{Tr}(\mathbf{Y}_{u} {\bf M}_{u}^{-1}) = \mbox{Tr}( \frac{\partial
  {\bf M}_u }{\partial v} {\bf M}_{u}^{-1}) = \frac{1}{\det{{\bf M}_u}}
\frac{\partial}{\partial v}\left( \det{{\bf M}_u}\right)
\eea
and for the down-type case
\bea
\sum_n \re\left({\frac{y^d_{nn}}{m^d_{n}}}\right)
=\mbox{Tr}(\mathbf{Y_{d} {M_{d}}^{-1}}) = \frac{1}{\det{\bf M_d}}
\frac{\partial}{\partial v}\left( \det{\bf M_d}\right)
\eea
We evaluate these traces perturbatively by expanding the
determinants in powers of $v^2/M_{KK}$ where
$M_{KK}$ are the masses of the heavy KK fermion excitations. 
We obtain
\bea
c^S_{ggh}
 &\simeq& v\  \re \Big[ 2\ \mbox{Tr}(\Delta^u_H+\Delta^d_H)
  +   \mbox{Tr}({y^0_d}^{-1} \Delta^u_2 + {y^0_d}^{-1}  \Delta^d_2 )\Big]\non\\ 
&&  \ \ +  \re \left( \frac{y_t}{m_{t}}\right) A^S_{1/2}(\tau_t)  +
\re \left( \frac{y_{b}}{m_{b}}\right) A^S_{1/2}(\tau_b) 
\eea
and
\bea
c^P_{ggh} &\simeq& v\ \im \Big[ \Big( 2\  \mbox{Tr}(\Delta^u_H+\Delta^d_H)
+\mbox{Tr}({y^0_u}^{-1}  \Delta^u_2 + {y^0_d}^{-1}  \Delta^d_2) \Big]\non\\
&&  \ \ +   \im \left( \frac{y_t}{m_{t}}\right) A^P_{1/2}(\tau_t)  +
  \im \left( \frac{y_{b}}{m_{b}}\right) A^P_{1/2}(\tau_b)
\eea
where the top and bottom quarks have been treated separately so as to
to compute numerically their associated loop functions
$A^{S/P}_{1/2}(\tau_t)$ and $A^{S/P}_{1/2}(\tau_b)$, without assuming
any limiting value for them. We now have to evaluate numerically all the terms in the previous
expressions. 

\bit
\item The terms depending on $\displaystyle
\frac{y_t}{m_t}$ and on  $\displaystyle \frac{y_b}{m_b}$ (the ratio of
the physical top Yukawa to its mass, and similarly for the bottom) are
obtained numerically using Eq.~(\ref{yukawann}), and the
perturbative numerical solutions from Eqs.~(\ref{ql}), (\ref{ur}), (\ref{qr}) and (\ref{ul}).

\item The terms $\mbox{Tr}({y^0_u}^{-1}  \Delta^u_2 + {y^0_d}^{-1}
\Delta^d_2)$ represent the ``kinetic'' shift in Yukawa couplings and
are highly suppressed, except for the top and bottom quarks, so we
rewrite them as 
\bea
&&\mbox{Tr}({y^0_u}^{-1}  \Delta^u_2 + {y^0_d}^{-1} \Delta^d_2)\  =\  
\frac{ \Delta^{t}_2}{y^{t}}+\frac{ \Delta^{b}_2}{y^{b}}
\  =\ \frac{1}{v^2} \left(\frac{v^3 \Delta^{t}_2}{m_{t}}+\frac{ v^3\Delta^{b}_2}{m_{b}} \right)  \non
\\
&=& \frac{1}{v^2} \sum_{i=1}^3 \int dz  a(z)^4  \Big(u^{i \,t*}_L u_L^{
  i\,t} + q_R^{i \,t*}q_R^{i \,t} + d^{i \,b*}_L d_L^{i \,b} + q_R^{i \,b*}q_R^{i \,b} \Big) 
\eea
These terms can be also be obtained using the perturbative numerical
solutions from Eqs.~(\ref{ql}), (\ref{ur}), (\ref{qr}) and (\ref{ql}).

\item Finally, the terms with $\Delta^u_H$ and $\Delta^d_H$, can be
calculated as
\bea
&&\mbox{Tr}(\Delta^u_H) =   \mbox{Tr}(M_Q^{-1} Y_1M_U^{-1} Y_2)\non \\
&&= \sum_{i,j =1}^3 \frac{Y^{5D}_{ij}}{\sqrt{k}} \frac{Y^{5D^*}_{ij}}{\sqrt{k}} 
\hspace{-.2cm} \int dy dy' e^{-4[A(y)+A(y')] } h(y)h(y')
\left( \sum_{n_i\ge1}^\infty   \frac{ Q^{n_i}_L(y)
  Q^{n_i}_R(y')}{M_{Q_{n_i}}} \right) \!\!\! 
\left( \sum_{m_i\ge1}^\infty  \frac{ U^{m_j}_R(y)  U^{m_j}_L(y')
}{M_{U_{m_j}}}  \right)\ \ \ \ \ 
\eea 
with the submatrices $M_U$,  $Y_2$, $M_Q$ and $Y_1$ as defined in
(\ref{mumatrix}), and with the down-type term obtained similarly.
Here, the wave functions in capital letters $Q^{n_i}_L(y)$, $Q^{n_i}_R(y)$, $U^{n_i}_R(y)$ and
$U^{n_i}_L(y)$ correspond to the KK quarks in the the {\it gauge basis}, when
$v=0$ (i.e. before electroweak symmetry breaking). The masses $M_{Q_{n_i}}$ and $M_{U_{n_i}}$ are
their corresponding masses (again, before electroweak symmetry breaking). These terms with infinite sums
can be calculated numerically after using the closure relations (see for
example \cite{Frank:2013qma})
\bea\label{sumuf}
\sum_{n=1}^{\infty}{{Q}_L^{(n)}(y){Q}_R^{(n)}(y')\over M_{Q_n}} 
= -e^{Q_q(y')-Q_q(y)} \left[\theta(y'-y)-{\int_0^{y'}e^{A-2Q_q}\over\int_0^{y_1}e^{A-2Q_q} }\right],
\eea
\bea\label{sumqf}
\sum_{n=1}^{\infty}{{U}_R^{(n)}(y){U}_L^{(n)}(y')\over m_n} = 
e^{Q_u(y)-Q_u(y')}\left[\theta(y'-y)-{\int_0^{y'}e^{A+2Q_u}\over\int_0^{y_1}e^{A+2Q_u} }\right].
\eea

\eit

In the case of the coupling of Higgs to photons, one proceeds in a
similar way in order to compute the fermion loops taking care to also add the
contribution from charged leptons and appropriately account for the
different gauge charges and coupling constant.

\section{Conclusions}
\label{sec:summary}

In this work we analyzed the production and decay rates of a
bulk-localized Higgs boson in the context of a general 5-dimensional
warped spacetime. Our analysis is concentrated on models with 
metrics modified from the usual RS model, which can account for low energy
electroweak and flavor precision measurements. These models
generically predict an enhancement in the Higgs production cross
section and a suppression in the fermion Yukawa couplings. Nevertheless
these predictions can remain in agreement with the LHC data, while
still allowing for a low KK scale,  within the LHC Run II reach.  In a
previous work, we presented an analysis  of the Higgs boson production which employed a toy model based
on one fermion field propagating in the bulk. We expand our considerations to present a more realistic analysis, which
 exhibits several new features.  

First, our investigations include a careful analysis to account for
all three families of quarks and leptons, together with their KK
towers. In particular, we start with fermion profiles which satisfy
masses and mixing constraints (as given by the CKM and PMNS matrices,
respectively).  We proceed to perform the calculation for production
and decay rates, first by considering the theory an effective theory
with a cut-off of  2 KK, then 3 KK modes, to test the convergence
properties. Second, we then compare the results with those obtained by
including a full  tower of KK modes for all fermion families. Careful
inclusion of the infinite KK tower, without neglecting the flavor
mixing, entails some technicalities,  described here in some
detail. Third, we include also the di-photon decay, which was not
evaluated in our previous toy model, and is calculated in the same way: first by
taking into account 2 KK and 3 KK modes, then including the full KK
tower.  And finally, we compare the results for our model with
modified metric to the RS limit (meaning that we assume the same
parameters for the RS limit, for a fair comparison, while a realistic
evaluation of RS models would have to take into account the fact that
the KK scale must be much higher).  

Our results are showcased as a function of the Higgs localization
parameter. For de-localized Higgs bosons (small localization
parameter, indicating a bulk Higgs), results obtained using 2 or 3 KK
modes agree with each other and with the infinite sum. Localizing the
Higgs closer to the brane enhances the gluon fusion cross section and worsens
the agreement, meaning that more KK modes are required for
agreement. We have chosen two ranges of values for the 5-dimensional Yukawa
coupling: $Y^{5D} \sim 1$ and $Y^{5D} \sim 3$. The later illustrates
the disagreement between including finite KK levels and including the
full towers, as one quickly reaches the non-perturbativity limit for our chosen (low) KK
masses. The behavior of $h \to \gamma \gamma$ has similar features to
the gluon fusion production: note that, as expected, the gluon fusion cross
section is enhanced, while the di-photon decay is suppressed
throughout the parameter space, confirming previous results from the one
fermion analysis. 

Our analysis is presented with the expectation that the Run II of the
LHC will measure Higgs boson properties with a high degree of
precision, and as such can put further limits on the parameters of
this model, or in fact rule it out. To compare with expected
measurements, we have included the top, bottom and tau lepton Yukawa
couplings  compared with the SM ones. We have chosen the warped
model with modified metric as the most promising model of extra
dimensions with consequences at low (within LHC reach) scales,
preferring it over competing models which include extra custodial
symmetries with additional fermions, gauge bosons and Higgs representations. A
careful and comprehensive analysis, such as this one, is timely and
can serve as a map towards revealing physics beyond the Standard
Model.

\section{Acknowledgements}
We thank NSERC and FRQNT for partial financial support under grant
numbers SAP105354 and PRCC-191578.



\begin{thebibliography}{99}
\bibitem{Aad:2015gba} 
 G.~Aad {\it et al.} [ATLAS Collaboration],
  Phys.\ Lett.\ B {\bf 716}, 1 (2012)
  [arXiv:1207.7214 [hep-ex]].

\bibitem{Chatrchyan:2012xdj} 
  S.~Chatrchyan {\it et al.} [CMS Collaboration],
  Phys.\ Lett.\ B {\bf 716}, 30 (2012)
  [arXiv:1207.7235 [hep-ex]].

\bibitem{Randall:1999vf} 
  L.~Randall and R.~Sundrum,
  Phys.\ Rev.\ Lett.\  {\bf 83}, 4690 (1999)
  [hep-th/9906064].
  L.~Randall and R.~Sundrum,
  Phys.\ Rev.\ Lett.\  {\bf 83}, 3370 (1999)
  [hep-ph/9905221].

\bibitem{stabilization}
W.~D.~Goldberger and M.~B.~Wise,
  Phys.\ Rev.\ Lett.\  {\bf 83}, 4922 (1999);
  W.~D.~Goldberger and M.~B.~Wise,
  Phys.\ Lett.\ B {\bf 475}, 275 (2000);
  C.~Csaki, J.~Erlich, T.~J.~Hollowood and Y.~Shirman,
  Nucl.\ Phys.\ B {\bf 581}, 309 (2000);
  C.~Csaki, M.~Graesser, L.~Randall and J.~Terning,
  Phys.\ Rev.\ D {\bf 62}, 045015 (2000);
  O.~DeWolfe, D.~Z.~Freedman, S.~S.~Gubser and A.~Karch,
  Phys.\ Rev.\ D {\bf 62}, 046008 (2000);
  N.~Arkani-Hamed, S.~Dimopoulos and J.~March-Russell,
  Phys.\ Rev.\ D {\bf 63}, 064020 (2001).
Also see: 
M.~Geller, S.~Bar-Shalom and A.~Soni,
  Phys.\ Rev.\ D {\bf 89}, no. 9, 095015 (2014). 
  
\bibitem{Davoudiasl:1999tf} 
  H.~Davoudiasl, J.~L.~Hewett and T.~G.~Rizzo,
  Phys.\ Lett.\ B {\bf 473}, 43 (2000); 
  S.~Chang, J.~Hisano, H.~Nakano, N.~Okada and M.~Yamaguchi,
  Phys.\ Rev.\ D {\bf 62}, 084025 (2000); 
  T.~Gherghetta and A.~Pomarol,
  Nucl.\ Phys.\ B {\bf 586}, 141 (2000).


 
\bibitem{Pomarol:1999ad} 
  A.~Pomarol,
  Phys.\ Lett.\ B {\bf 486}, 153 (2000).
  

\bibitem{Grossman:1999ra} 
  Y.~Grossman and M.~Neubert,
  Phys.\ Lett.\ B {\bf 474}, 361 (2000).
  

\bibitem{Chang:1999nh} 
  S.~Chang, J.~Hisano, H.~Nakano, N.~Okada and M.~Yamaguchi,
  Phys.\ Rev.\ D {\bf 62}, 084025 (2000).
  

\bibitem{Gherghetta:2000qt} 
  T.~Gherghetta and A.~Pomarol,
  Nucl.\ Phys.\ B {\bf 586}, 141 (2000).
  

\bibitem{Davoudiasl:2000wi} 
  H.~Davoudiasl, J.~L.~Hewett and T.~G.~Rizzo,
  Phys.\ Rev.\ D {\bf 63}, 075004 (2001).


\bibitem{Contino:2003ve} 
  R.~Contino, Y.~Nomura and A.~Pomarol,
  Nucl.\ Phys.\ B {\bf 671}, 148 (2003)
  [hep-ph/0306259].
 
\bibitem{Csaki:2008zd} 
  C.~Csaki, A.~Falkowski and A.~Weiler,
  JHEP {\bf 0809}, 008 (2008)
  [arXiv:0804.1954 [hep-ph]].
 

\bibitem{Carena:2004zn} 
  M.~S.~Carena, A.~Delgado, E.~Ponton, T.~M.~P.~Tait and C.~E.~M.~Wagner,
Phys.\ Rev.\ D {\bf 71}, 015010 (2005);
  M.~S.~Carena, A.~Delgado, E.~Ponton, T.~M.~P.~Tait and C.~E.~M.~Wagner,
Phys.\ Rev.\ D {\bf 68}, 035010 (2003);
  A.~Delgado and A.~Falkowski,
JHEP {\bf 0705}, 097 (2007).


\bibitem{Csaki:2008qq} 
  C.~Csaki, C.~Delaunay, C.~Grojean and Y.~Grossman,
  JHEP {\bf 0810}, 055 (2008).


\bibitem{Agashe:2006at} 
  K.~Agashe, R.~Contino, L.~Da Rold and A.~Pomarol,
  Phys.\ Lett.\ B {\bf 641}, 62 (2006).
  

\bibitem{Huber:2003tu} 
  S.~J.~Huber,
  Nucl.\ Phys.\ B {\bf 666}, 269 (2003).
  
  
\bibitem{Agashe:2004cp} 
  K.~Agashe, G.~Perez and A.~Soni,
  Phys.\ Rev.\ D {\bf 71}, 016002 (2005);
  K.~Agashe, A.~Delgado, M.~J.~May and R.~Sundrum
  
    
\bibitem{Agashe:2003zs}
  K.~Agashe, A.~Delgado, M.~J.~May and R.~Sundrum,
  JHEP {\bf 0308} (2003) 050.
Also see: 
  B.~M.~Dillon and S.~J.~Huber,
  JHEP {\bf 1506}, 066 (2015)
  
\bibitem{Falkowski} 
  A.~Falkowski and M.~Perez-Victoria,
  JHEP {\bf 0812}, 107 (2008).


\bibitem{Batell} 
  B.~Batell, T.~Gherghetta and D.~Sword,
  Phys.\ Rev.\ D {\bf 78}, 116011 (2008);
  T.~Gherghetta and D.~Sword,
  Phys.\ Rev.\ D {\bf 80}, 065015 (2009);
  T.~Gherghetta and N.~Setzer,
  Phys.\ Rev.\ D {\bf 82}, 075009 (2010).





  

\bibitem{Cabrer:2011fb} 
  J.~A.~Cabrer, G.~von Gersdorff and M.~Quiros,
  JHEP {\bf 1105}, 083 (2011);
  J.~A.~Cabrer, G.~von Gersdorff and M.~Quiros,
  Phys.\ Lett.\ B {\bf 697}, 208 (2011);
  J.~A.~Cabrer, G.~von Gersdorff and M.~Quiros,
  New J.\ Phys.\  {\bf 12}, 075012 (2010);
  J.~A.~Cabrer, G.~von Gersdorff and M.~Quiros,
  Phys.\ Rev.\ D {\bf 84}, 035024 (2011);
  J.~A.~Cabrer, G.~von Gersdorff and M.~Quiros,
  JHEP {\bf 1201}, 033 (2012).



  
\bibitem{Carmona:2011ib} 
  A.~Carmona, E.~Ponton and J.~Santiago,
  JHEP {\bf 1110}, 137 (2011).



\bibitem{MertAybat:2009mk} 
  S.~Mert Aybat and J.~Santiago,
Phys.\ Rev.\ D {\bf 80}, 035005 (2009);
  A.~Delgado and D.~Diego,
Phys.\ Rev.\ D {\bf 80}, 024030 (2009).
  





  
\bibitem{Frank:2013un} 
  M.~Frank, N.~Pourtolami and M.~Toharia,
 Phys.\ Rev.\ D {\bf 87},\ 096003 (2013).
  
\bibitem{Frank:2013qma} 
  M.~Frank, N.~Pourtolami and M.~Toharia,
  Phys.\ Rev.\ D {\bf 89}, no. 1, 016012 (2014).
  
\bibitem{Archer:2014jca} 
  P.~R.~Archer, M.~Carena, A.~Carmona and M.~Neubert,
  JHEP {\bf 1501}, 060 (2015).
  Also see: 
  U.~K.~Dey and T.~S.~Ray,
  Phys.\ Rev.\ D {\bf 93}, no. 1, 011901 (2016)
  
 \bibitem{Azatov:2010pf} 
  A.~Azatov, M.~Toharia and L.~Zhu,
  Phys.\ Rev.\ D {\bf 82}, 056004 (2010); 
  A.~Azatov, M.~Toharia and L.~Zhu,
  Phys.\ Rev.\ D {\bf 80}, 035016 (2009).
 
\bibitem{Agashe:2014kda} 
  K.~A.~Olive {\it et al.}  [Particle Data Group Collaboration],
  Chin.\ Phys.\ C {\bf 38}, 090001 (2014).

%

\bibitem{Huber:2000fh}
  S.~J.~Huber and Q.~Shafi,
  Phys.\ Rev.\ D {\bf 63} (2001) 045010; 
  S.~J.~Huber, C.~A.~Lee and Q.~Shafi,
  Phys.\ Lett.\ B {\bf 531}, 112 (2002); 
  C.~Csaki, J.~Erlich and J.~Terning,
  Phys.\ Rev.\ D {\bf 66}, 064021 (2002); 
  J.~L.~Hewett, F.~J.~Petriello and T.~G.~Rizzo,
  JHEP {\bf 0209}, 030 (2002); 
  G.~Burdman,
  Phys.\ Rev.\ D {\bf 66}, 076003 (2002).

\bibitem{Gunion:1989we} 
  J.~F.~Gunion, H.~E.~Haber, G.~L.~Kane and S.~Dawson,
  Front.\ Phys.\  {\bf 80}, 1 (2000).

  
\bibitem{HiggsReview} A.~Djouadi, Phys. Rept. \textbf{457} (2008) 1,
 A.~Djouadi, Phys. Rept. \textbf{459} (2008) 1,
 J. Gunion, H. Haber, G. Kane, and S. Dawson, \emph{The Higgs Hunter's Guide} (Westview press, Boulder, CO, 2000). 
 
%
%
\bibitem{Bouchart:2009vq} 
  C.~Bouchart and G.~Moreau,
  Phys.\ Rev.\ D {\bf 80}, 095022 (2009)
  [arXiv:0909.4812 [hep-ph]].
  
%
%
\end{thebibliography}
\end{document}